\documentclass[preprint2]{aastex}

\usepackage[english]{babel}
\usepackage[utf8]{inputenc}
\usepackage{amsmath}
\usepackage{graphicx}
\usepackage[colorinlistoftodos]{todonotes}

\newcommand{\rr}[1]{\ \mathrm{#1}}

\slugcomment{Accepted by \textit{The Astronomical Journal}}
\shorttitle{Multiplexing Precision RVs}
\shortauthors{Bailey et al.}

\begin{document}

\title{Multiplexing Precision RVs: Searching for Close in Gas Giants in Open Clusters}

\author{John I. Bailey, III\altaffilmark{1}, Mario Mateo}
\affil{Department of Astronomy, University of Michigan}
\affil{1085 South University, Ann Arbor, MI  48109, USA}
\altaffiltext{1}{Current Affiliation: Leiden Observatory, Leiden University, P.O. Box 9513, 2300RA Leiden, The Netherlands}
\email{baileyji@umich.edu}
\author{Russel J. White}
\affil{Department of Physics \& Astronomy, Georgia State University}
\affil{P.O. Box 4106, Atlanta, GA 30302, USA}
\author{Stephen A. Shectman, Jeffrey D. Crane} 
\affil{Carnegie Observatories}
\affil{813 Santa Barbara Street, Pasadena, CA 91101, USA}
\and
\author{Edward W. Olszewski} 
\affil{Steward Observatory}
\affil{933 North Cherry Avenue, Tucson, AZ  85721, USA}


\begin{abstract}
We present a multiplexed, high-resolution (R$\sim$50,000 median) spectroscopic survey designed to detect exoplanet 
candidates in two southern star clusters (NGC~2516 and NGC~2422) using the Michigan/Magellan Fiber System (M2FS) 
on the Magellan/Clay telescope at Las Campanas Observatory. With 128 available fibers in our observing mode, we are
able to target every star in the core half-degree of each cluster that could plausibly be a solar-analog member. 
Our template-based spectral fits provide precise measurements of fundamental stellar properties---$T_{eff}$ ($\pm30$~K), 
[Fe/H] and [$\alpha$/Fe] ($\pm0.02$~dex), and $v_r\sin(i)$ ($\pm$0.3~km/s)---and radial velocities (RVs) by using telluric 
absorption features from 7160 to 7290~\AA\ as a wavelength reference for 251 mid-F to mid-K stars (126 in 
NGC~2516 and 125 in NGC~2422) that comprise our survey. In each cluster we have obtained $\sim10-12$ epochs of our targets. Using repeat observations of an RV standard star we show our approach can attain a single-epoch velocity precision of 
25~m/s to 60~m/s over a broad range of S/N throughout our observational baseline of 1.1~years. Our technique is suitable for 
non-rapidly rotating stars cooler than mid-F. In this paper we describe our observational sample, analysis methodology, and 
present a detailed study of the attainable precision and measurement capabilities of our approach. Subsequent papers 
will provide results for stars observed in the target clusters, analyze our dataset of RV time-series for stellar 
jitter and stellar and sub-stellar companions, and consider the implications of our findings on the clusters themselves. 
\end{abstract}


\maketitle

\section{Introduction}
The realization that planetary systems can harbor hot gas giants led to drastic revisions to both theories of planet formation and dynamical evolution. Contemporary formation models typically fall into the broad categories of core accretion \citep{mizuno80} or disk instability models \citep{boss97}, neither of which has yet emerged as a dominate driver. It also remains unclear if the orbits of the so-called Hot-Jupiters evolve inward through coupling to a gaseous disk \citep{goldreichtremaine80,lin96}, via dynamical scattering off clumps and other bodies \citep[e.g.,][]{rasio96,juric08}, or even secular interactions with a distant stellar companion \citep[Kozai cycles; e.g.,][]{fabrycky07}. Though much effort has been spent testing these theories they all remain viable, at least for some systems \citep[see review by][]{helled14}. The core reason is that diagnostics (e.g. the period-eccentricity distribution) are muddled by the wide and often unknown ages of host stars in field star exoplanetary systems. A direct solution would be to find young exoplanets with ages $\sim$1~Myr, however the extreme activity of host stars at T Tauri ages has thus far inhibited detecting planets (e.g. \citealp{2008ApJ...678..472H}; \citealp{2008A&A...489L...9H}; \citealp{2008ApJ...687L.103P}; \citealp{2012ApJ...761..164C}; \citealp{ysasref}); some very young candidates nevertheless remain \citep[e.g.][]{2015ApJ...809...42C}.

These theories do make various distinct predictions regarding the properties of hot gas giants and their orbits for the first $\sim1$ Gyr \citetext{c.f. \citealp{adams06,marley07,fortney10,fortney08} vs. \citealp{galvagni12}; but see also \citealp{mordasini12}}, and though some constraints require direct measures of properties via transits, which are rare, they are relatively more likely for hot gas giants \citep[$\sim$ 5\% for hot gas giants;][]{charbonneau07}. Indeed, considerable effort has been expended searching for such systems with known ages \citep[e.g.,][]{paulson04,ysasref}. The first exoplanets orbiting solar-like stars in a cluster (F, G, or K and on the main sequence) were announced by \citet{quinn12}. To date, gas giant exoplanets orbiting main sequence (MS) stars in clusters have been discovered via transits \citep[NGC 6811][]{meibom13} and RV techniques \citetext{M 67; \citealp{brucalassi14}; Praesepe, \citealp{quinn12}; Hyades, \citealp{quinn14}}, though none of the latter are also transiting systems.  One system discovered in the Hyades is noteworthy for its distinctly non-zero eccentricity ($e=0.08\pm0.02$), which implies that dynamical scattering has likely played a role in its dynamical evolution, at least in the late stages of the process \citep{quinn14}. It is already clear that even a few hot gas giant exoplanets in open clusters with precise ages have the potential to strongly constrain planet-migration theories.  Any such systems that also happen to be transiting will produce even more powerful constraints on gas giant formation models by revealing precise information on the sizes, densities and compositions of exoplanets with well-determined ages. Indeed, identifying transiting planets in open clusters is a key science goal of NASA's K2 mission \citep{howell14}.

The practical problem of finding cluster exoplanets is one of efficiency. Over the two decades of exoplanetary searches, 
more than 600 systems have been  studied using RV techniques. In contrast, there are only eight known exoplanets orbiting MS 
stars in open clusters.  The deficiency of known exoplanets in clusters is not a consequence of higher stellar RV jitter.  Measured 
values of this in open clusters are $\sim15$ m/s at $\sim400$ Myr \citep{paulson04,quinn12}, which is roughly an order of 
magnitude less than the amplitude induced by a typical hot gas giant. Rather, clusters tend to be considerably more distant than 
the closest individual field stars and are hence fainter and consequently harder to monitor.  Moreover the same technique used 
for field stars -- individual spectra, one star at a time -- is typical of cluster surveys. Lengthy campaigns involving thousands of 
visits are needed to find the comparatively rare cases with detectable velocity amplitudes: only $1.2 \pm 0.4$\% of all FGK stars 
in the solar neighborhood harbor hot gas giants \citep{wright12} and cluster occurrence rates appear similar \citep{meibom13} so 
hundreds of targets must be monitored. The MARVELS survey \citep{marvels} is a notable exception to this with its $\sim$11,000 
stars and is intended to attain similar RV precision to the work we report here though it is not targeted to clusters. 

In this paper we introduce a different approach that employs highly-multiplexed RV techniques to detect and study exoplanetary 
systems containing hot gas giants down to $\sim0.1\rr{M}_{Jup}$ that are orbiting stars in clusters that range in age from about 
100 Myr to nearly 1 Gyr. To our knowledge our survey is the first multi-object spectroscopic survey of open cluster stars to obtain RV precisions at the sub-hundred m/s level. To do this, we have used spectra obtained with the Michigan/Magellan Fiber System 
\citep[M2FS;][]{m2fsspie} which allows us to obtain multiplexed, high-resolution (R $\sim50,000$) optical spectra of 
solar-analogue stars (spectra types F5V to K5V) in nearby ($\lesssim$1~kpc) clusters. Our approach -- a variant of the telluric-reference approach first proposed by \citet{griffin} and subsequently used or studied by \citet[][and references therein]{cochran88}, \citet{seifahrt10}, and \citet{ysasref} -- models the observed stellar spectra and telluric absorption features to obtain high-precision velocities and stellar parameters. As part of this process we map out the RV precision attainable using telluric lines across a wide range of wavelengths and quantify the assertions of \citet{griffin} from some 43 years ago.
We are able to measure RVs to 25 m/s for a slowly rotating, bright RV standard star and 30--60~m/s for up to 128 stars simultaneously over a half-degree field-of-view. We expect our precision to remain better than $\sim$75~m/s for stars as faint as $V=17$. Though our precision is not particularly notable when compared with e.g. the 0.3~m/s precision of HARPS \citep{harps} or similar instruments, it is comparable to the expected performance of MARVELS and our magnitude limit is significantly beyond the survey's V=12 limit. HARPS is capable of 30~m/s precision at $V=16.6$ but it requires $\sim$1~hour exposure -- 4000 times less efficient than our multi-object approach. 

The goal of this paper is to describe the details of our methodology. We will address the stellar properties and RV variability of our target stars in future publications. We describe our target selection and observing procedure in Section~\ref{observations}. Section~\ref{reductionsec} describes our image reduction procedure and extraction to 1D spectra. In Section~\ref{analysis} we present our spectral modelingapproach and describe the procedure in detail. Section~\ref{resultssec} analyzes the  quality of the measured RVs and stellar parameters. Finally Section~\ref{discussionsec} considers the implications of our results for the future of this survey technique.

\section{Observations}
\label{observations}

\subsection{Cluster and Spectroscopic Target Selection}
\label{targselecsec}

\begin{deluxetable}{lrl}
\tablewidth{0pc}
\tablecaption{Cluster Selection Criteria\label{selecttable}}
\tablehead{
\colhead{Criteria} &
\colhead{Value} &
\colhead{Comment} 
}
\startdata
Dec (deg) & $<+10$ & Visible at Magellan \\
DM & 10.0 & Bright enough for pRVs \\
Age (Myr) & $\gtrsim100$ & Limit stellar activity \\
$\rr{R_{cen}}$ (deg) & $\lesssim1.0$ & Match to M2FS FOV \\
Fe/H & $\gtrsim-0.3$ & Enhance HJ formation \\
$\rr{N_{F5-K5}}$ & $\gtrsim80$ & Match to number of fibers\\
\enddata
\end{deluxetable}

To select our targets, we first created a list of potential star clusters in the Catalog of Open Cluster Data \citep{khar05ref} suitable for this project using a small number of basic selection criteria (Table~\ref{selecttable}) chosen to identify systems suitable for multi-object spectroscopy of solar analogs that are close enough to have members sufficiently bright to detect small-amplitude RV variability and old enough to limit stellar jitter. We imposed restrictions on cluster size ($\mathrm{R_{cen}}$ in Table~\ref{selecttable}, the core radius derived by \citealp{khar05ref}) and richness to ensure good multiplexing efficiency, and placed limits on age and metallicity to exclude clusters with stars that exhibit excessive surface activity (thereby mimicking or masking the Doppler RV variations of a companion) and increase the likelihood of gas giant planet formation. From the list of 124 clusters meeting the first four criteria in Table~\ref{selecttable} we identified  $\sim$30 matching clusters. Ultimately we selected NGC~2516 and NGC~2422 as our targets as they were the richest clusters observable during our first observing run in November 2013. These $\sim$140~Myr and $\sim$75~Myr old open clusters are within 500 pc, rich in solar analogues, have angular sizes that are well-matched to the multiplexing capabilities of M2FS, and have recent photometric membership catalogs sufficiently deep for selecting solar-analog members \citep[][hereafter J01 and P03]{n2516ref, n2422ref}.

Individual targets in NGC~2516 were drawn from the sample of stars studied in J01. We selected all stars they identified as photometric single (79) or photometric binary (47) members having colors and magnitudes consistent with F5V--K5V spectral types. This sample of 126 stars was then cross-matched with the UCAC4 catalog \citep{ucac4} and the UCAC4 coordinates used to prepare the plug plate. In NGC~2422 we selected all objects with colors and magnitudes consistent with F5V--K5V in the membership list of P03. Due to a smaller number of members in our field of view, we expanded our selection out in color from the main sequence defined by P03 members using the UCAC4 catalog until we had sufficient targets to fill the available fibers, eventually selecting an additional 25 stars in our adopted pointing in this manner. We then cross-matched the P03 targets with UCAC4 for astrometry. With 128 available fibers, we are able to target every star in each half-degree field that could plausibly be a solar-analog member. Table~\ref{targtable1} provides a summary of the clusters, our adopted pointings, and the stellar targets therein.

We selected an additional field (also within NGC~2516) with thirty-two sources with magnitudes (assuming membership) and 
colors consistent with B8-A4 stars (\bv$=-0.1$ -- 0.113; $\rr{M_V}=0$ -- 1.7). These stars possess essentially featureless spectra -- apart from telluric features -- in our wavelength range and hence can serve as useful probes to monitor the instrumental point spread function (PSF) over the full field of view of the spectrograph cameras.

Finally, we selected six stars with similar RAs from the GAIA RVS catalog \citep{gaiarv} for use as radial velocity standards. 
One of these -- HIP 48331 -- was observed repeatedly and is used as a primary reference to track our RV measurement 
precision over the duration of the program. Although this star hosts an exoplanetary companion, the induced RV semi-amplitude 
is only 0.8 m/s \citep{hip48331planet}, far below our measurement precision and so 
its variability is irrelevant for our purposes.  A summary of the standard stars used for this study is presented in Table~\ref{stdtargtable}.

\begin{deluxetable}{lccc ccc cccc}
\tablewidth{0pc}
\tabletypesize{\scriptsize}
\setlength{\tabcolsep}{0.03in}
\tablecaption{Target Cluster and Pointing Information\label{targtable1}}
\tablehead{
\colhead{} & \colhead{}  &
\colhead{RA} & \colhead{Dec.} & \colhead{Age} & \colhead{Dist.} & 
\colhead{} & \colhead{}  & \colhead{}  & \colhead{}  &\colhead{} 
\\
\colhead{Cluster} & \colhead{Messier} &
\colhead{(2000)} & \colhead{(2000)} & \colhead{(Myr)} &  \colhead{(pc)} &
\colhead{E(B$-$V)} & \colhead{$\rr{N}_{ep}$} & \colhead{$\rr{N}_{targ}$}& \colhead{V} &
\colhead{B$-$V}
}
\startdata
NGC~2516 &... & 7:58:42 & $-$60:46:36 & 141 & 346 & 0.11 & 12 & 126 &11.68--15.09 & 0.46--1.26 \\
NGC~2422 & M 47 & 7:36:30 & $-$14:29:42  & 72 & 491 & 0.07 & 10 & 125 & 12.20--16.10 & 0.45--1.43 \\
\enddata
\tablecomments{The coordinates listed correspond to our field centers and, although near, are not at 
the cluster center. Both distances and the reddening value for NGC~2422 are from \citet{khar05ref}. 
Target photometry is from J01 (NGC~2516) and P03/UCAC4 (NGC~2422). The reddening value for 
NGC~2516 is taken from \citet{n2516sung}. The age for NGC~2516 is from \citet{n2516age} and for NGC~2422 from 
\citet{n2422youngage}. We note that \citet{khar05ref} gives ages of 120~Myr NGC~2516 and 132~Myr for NGC~2422, 
albeit with errors $\sim$70~Myr.}
\end{deluxetable}

\begin{deluxetable}{lllrcccccc}
\tablewidth{0pc}
\tabletypesize{\scriptsize}
\tablecaption{Standard Stars\label{stdtargtable}}
\tablehead
{
\colhead{} & 
\colhead{} & 
\colhead{} &
\colhead{RV } &
\colhead{$v \sin(i)$} & 
\colhead{}  & 
\colhead{$T_{eff}$} & 
\colhead{} & 
\colhead{} & 
\colhead{}
\\
\colhead{Target} & 
\colhead{N} & 
\colhead{$V_{mag}$} &
\colhead{(km/s)} &
\colhead{(km/s)} & 
\colhead{Sp. Type}  & 
\colhead{(K)} & 
\colhead{log(g)} & 
\colhead{[Fe/H]} & 
\colhead{[$\alpha$/Fe]}}
\startdata

HIP 48331 &  35 & 7.67 & $-9.510\pm0.005 $ & 0.9 & K5V &  $4455\pm80$ & 4.67 & -0.18\tablenotemark{a}  & ...   \\
HIP 13388 &  2   & 8.09 & $65.606\pm0.009 $ & 2.7 & K1V & $5095\pm64$ & 4.59 & -0.15 & 0.02  \\
HIP 10798 &  5   & 6.33 & $ 7.469\pm0.007 $ & 2.7 & G8V & $5481\pm80$ & 4.63 & -0.44 & 0.17  \\
HIP 22278 &  3   & 8.52 & $ 23.456\pm0.014 $ & 3.6 &G5V & $5721\pm65$ & 4.22 & 0.13 & -0.01  \\
HIP 19589 &  1   & 8.46 & $-5.500\pm0.024 $ & 3.6 &G0V & $5825\pm90$  & 3.75 & -0.17 & 0.13  \\
HIP 31415 &  1   & 7.70 & $-7.479\pm0.012 $ & 4.5 &F6V & $6172\pm60$ & 3.94 & -0.31 & 0.12  \\

\enddata
\tablenotetext{a}{Taken from \citet{santos05} as \citet{Casagrande11} flags these measurements as of poor quality, though
Santos reports an error of  $0.19$. }
\tablecomments{This table lists the literature properties and number of epochs we obtained of the standard stars 
observed for this program. RVs, magnitudes, and spectral types are taken from \citet{gaiarv}. $v_r\sin(i)$ values are from \citet{Glebocki05}. 
Stellar parameters are taken from from \citet{Casagrande11}.}
\end{deluxetable}

\subsection{Instrument Configuration}
The Michigan/Magellan Fiber System (M2FS) is a multi-object, optical (3700 - 9500 \AA), fiber-fed spectrograph 
that can take simultaneous spectra of up to 256 targets over a half-degree field-of-view at a wide variety of resolutions
($R\sim500 - 55,000$). The fibers accept light from the sky at the f/11 Nasmyth E focal surface of the Magellan/Clay 
6.5 m telescope at Las Campanas Observatory.  Each fiber samples the sky through a 1.2 arcsecond aperture; fibers can be 
packed to within 14 arcseconds of one another with no restrictions within the M2FS field of view and are held in place by plug 
plates drilled in advance of an observing run using astrometry for the desired targets. 

\begin{figure*}
\plotone{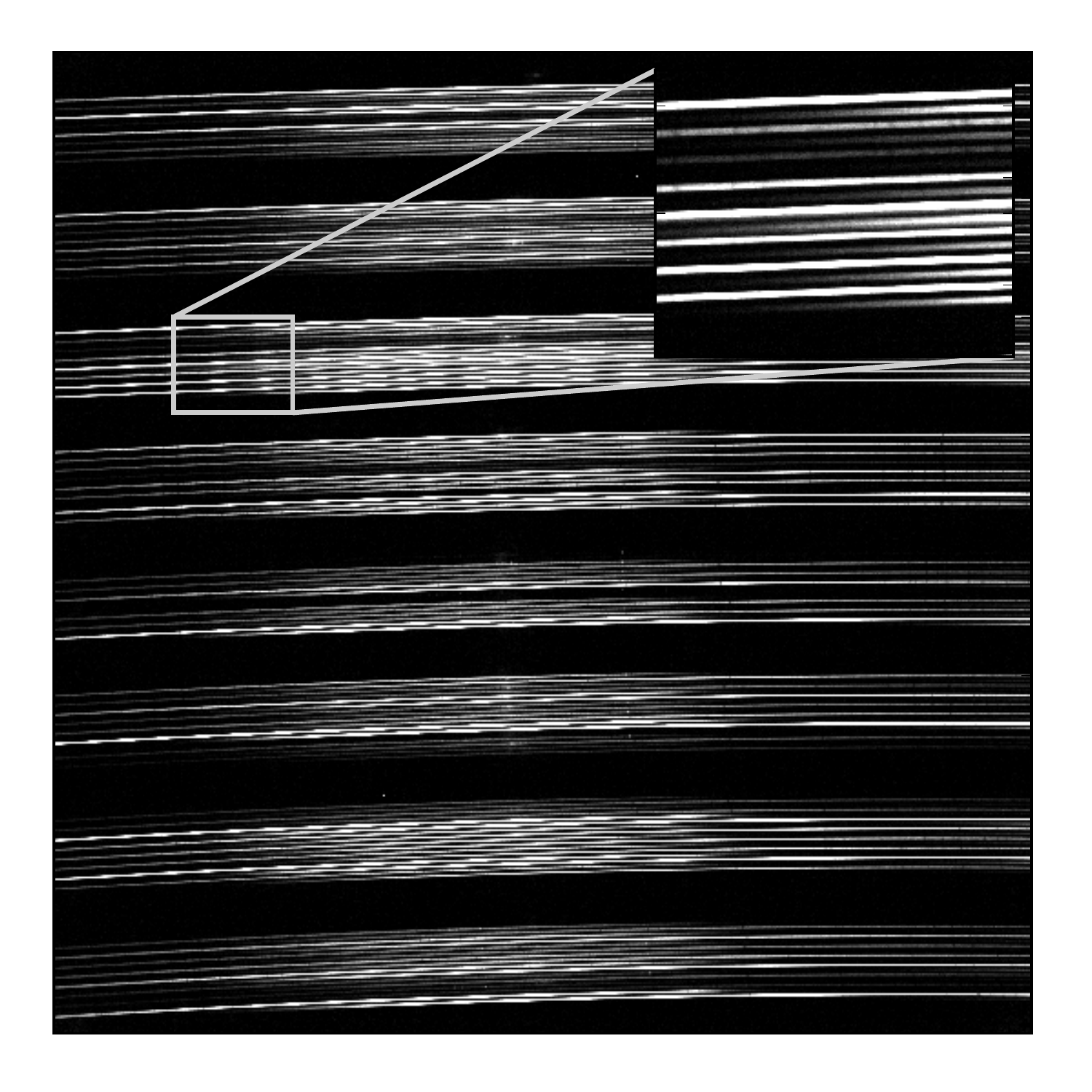}
\caption{\label{sampleim} An example science frame in NGC~2516. M2FS fibers are bundled in groups of sixteen at the camera focal plane, 
and although we use every other fiber we also use a two-order passband, resulting in the groups of sixteen spectra. The larger gaps in the image
reflect spacing between adjacent bundles of fibers and are used to estimate the scattered light in the image. 
Each fiber maps to a consecutive pair of apertures (best seen in the inset): the lower is order 50 (unused) and the upper is order 49. 
The variability in this frame is a function of both target magnitude and fiber throughput.}
\end{figure*}

The fibers terminate at the focal surface of the collimator/cameras belonging to a pair of identical quasi-Littrow
spectrographs. For historical reasons, these are identified as the `red' and `blue' M2FS arms, each fed by 128 fibers.
The spectrographs are equipped with traditional and echellette gratings for low and medium resolution work as
well as an R2.0 echelle grating with a prism cross-disperser for high-resolution use.  Filters located just below the 
fiber termination surface isolate specific orders, necessary for use in echelle and echellette modes. Each arm uses an 
E2V 4k x 4k anti-fringing CCD with 15 $\mu$m pixels. Additional details can be found in \citet{m2fsspie}. 
Figure~\ref{sampleim} shows an example frame from our program.

\begin{figure*}
\centering\includegraphics[width=.6\textwidth,clip=true,trim=0cm 0cm 0cm 0cm]{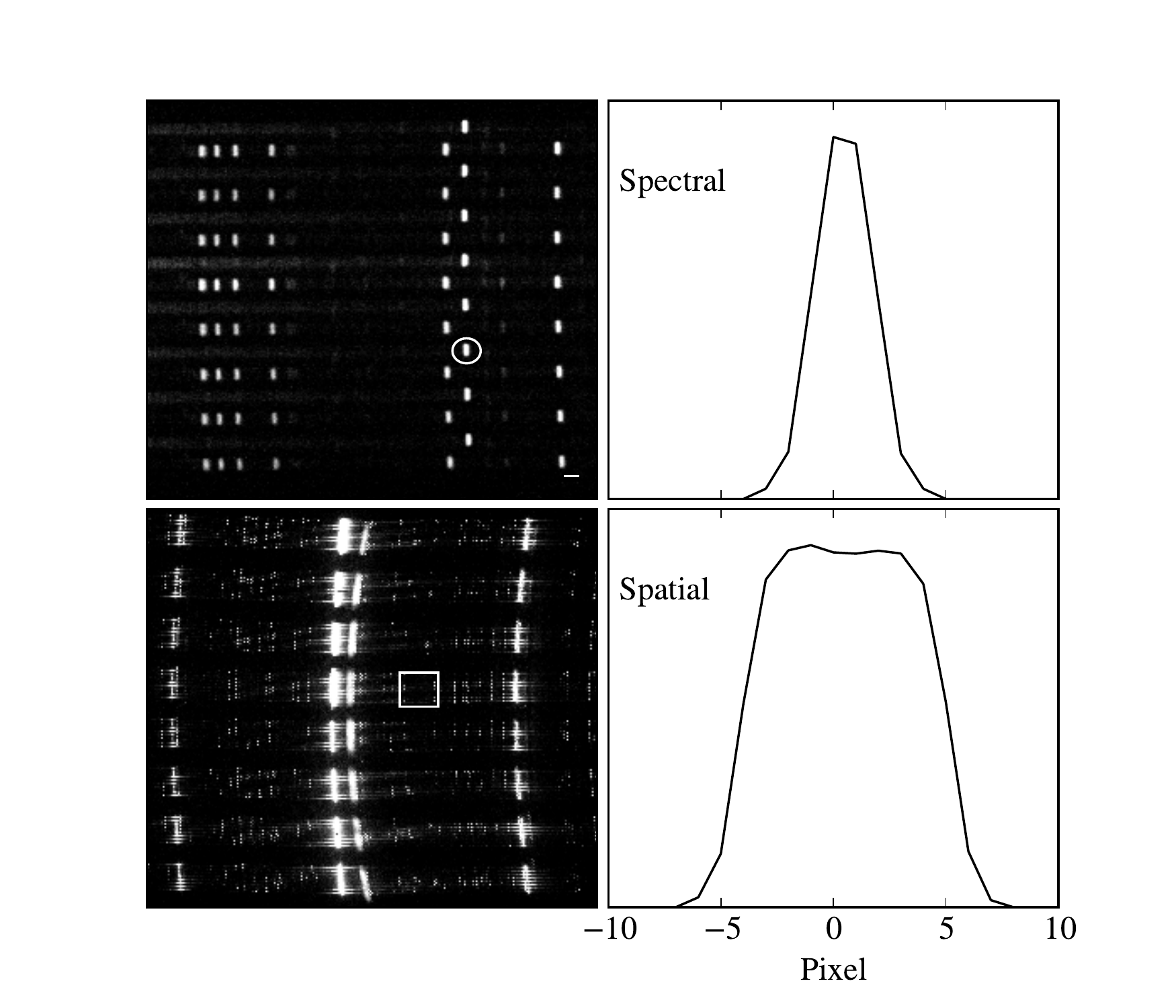}
\caption{\label{slitlampimage}
This figure shows various aspects of an M2FS ThArNe calibration image (lower-left) made in our 
configuration. The upper-left panel is detailed view of the boxed region. In it the small horizontal bar 
to the lower-right of the image is 10 pixels long and the line profiled in the right two panels is encircled. 
The FWHM of the lamp lines across the images corresponds to $\sim3$ pixels on 
average (R$\sim$50,000).
}
\end{figure*}

M2FS incorporates a slit mechanism placed just after the ends of the fibers at the spectrograph collimator/camera focal surface; 
details of this mechanism are provided in \citet{Bailey:2012ie}. For all of the HiRes observations used in this paper, we 
employed the 45-micron wide slit (the narrowest available) which projects to approximately three pixels at the detectors.  
For this configuration we found an effective resolution of in 40-60k range for most fibers (c.f Figure~\ref{slitlampimage}), subject to focus variations. 
Because the slits are positioned after the fibers we sacrifice roughly 60\% of the incident  light.

\begin{figure*}
\centering\includegraphics[width=.9\textwidth,clip=true,trim=2.5cm 0cm 4.5cm 0cm]{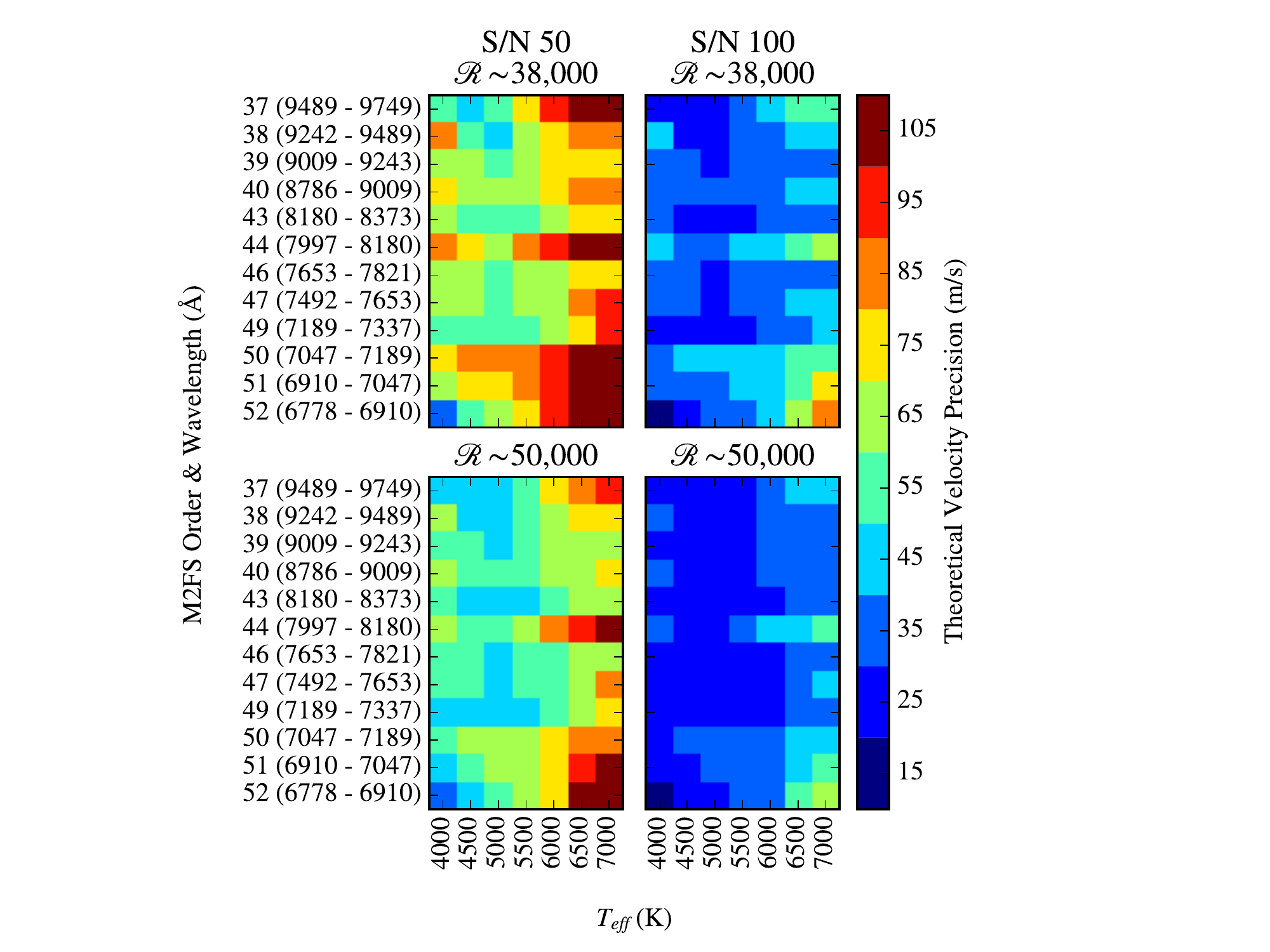}
\caption{\label{precisionfig}The velocity precision attainable for slowly rotating ($v_r\sin(i)=5$ km/s), Solar abundance,
dwarfs stars with $T_{eff}$ between 4000 K and 7000 K when using telluric lines as the wavelength reference, 
according to the formalism of \citet{attaining3ms}.
Each individual plot gives M2FS echelle order number and nominal wavelength range on the vertical axis
and $T_{eff}$ on the horizontal axis. In each case a log(g) of 4.5 is used, though results are not significantly 
affected by this choice. The two columns correspond to S/N of 50 and 100 and the rows to $\lambda/d\lambda$ of 
38,000 and 50,000. Colors correspond to the attainable RV precision. It is interesting to note that this plot 
quantifies the assertions of \citet{griffin} from some 43 years ago.}
\end{figure*}

For this study, we employed a two order filter with a design passband of 7050--7370~\AA\ 
(M2FS echelle orders 49 and 50): the use of a filter is necessary to prevent spectral orders from one fiber overlapping 
with those of another and a two order filter limits us to every second M2FS fiber for a multiplexing factor of 128. 
As few target clusters would offer more than about 100 targets in a single M2FS field this decision does not appreciably affect our multiplexing capability. The passband was selected after careful consideration of the optimal wavelength region to 
carry out telluric-reference RV measurements of solar analogue stars.  We used the formalism of \citet{attaining3ms} combined with synthetic spectra from the PHOENIX grid \citep{phoenix} to estimate the RV uncertainties for slowly-rotating, main-sequence stars with effective temperatures between 4000 and 7000 K for a range of M2FS orders red-ward of about 6800 \AA (where telluric features become common). This uncertainty was then added in quadrature with the wavelength reference uncertainty in each order determined by applying the same formalism to the telluric absorption features present in the empirical telluric spectrum from \citet{wallaceref}. Results of this analysis are shown in Figure~\ref{precisionfig}: we show our estimates of attainable velocity precision for Sun-like, slowly rotating stars observed at two fiducial resolving powers and S/N levels for each M2FS order with appreciable telluric lines. 

\begin{figure}
\plotone{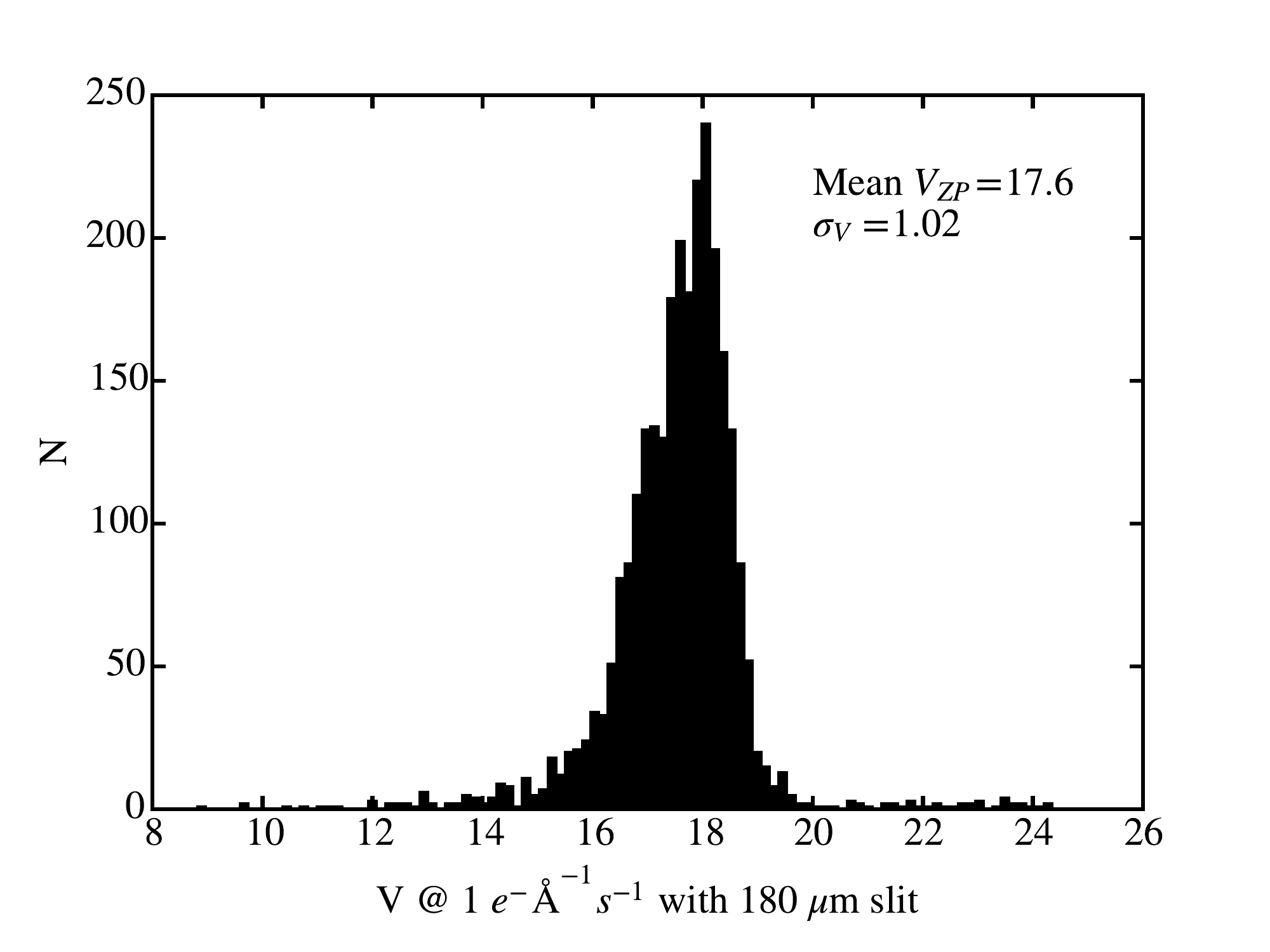}
\caption{\label{throughputfig} A plot of M2FS system throughput as calculated for each of our 2700 spectra. Values have been 
corrected for both variations in fiber throughput and for the throughput losses imposed by our use of the 45 
$\mu$m slit ($\sim60$\%) and assume our median seeing of $\sim0.8$\arcsec. For other M2FS instrument modes we find a typical zero 
point of $18.3 \pm 0.3$ mag, significantly fainter that measured for these spectra. This loss is a direct result of our filter 
bandpass edge falling just prior to the blaze peak. This is corrected in the second version of our filter.}
\end{figure}

In practice we only obtain useful data from the first portion of order 49 (7160--7290~\AA{}). Our filter was manufactured prior to
the discovery of a slight difference in the intended and manufactured echelle grating blaze. This rendered data from order 50 to 
be of little use, truncated the redder portion of order 49, and limited system throughput by a factor of two 
(see Figure~\ref{throughputfig}). This limited bandpass was still able to deliver excellent RV precision (Section~\ref{rvprecisionsec}), 
and manageable S/N for stars in our target clusters. In early 2016 we started verification tests on a new filter designed to 
cover the 7160--7360~\AA{} region, adding an additional 14 major lines and recovering the lost throughput. We anticipate this 
will improve our achievable velocity precision by $\sim$15\% where we are not limited by systematics and do not anticipate 
any negative impact on our program. 
\label{filterissuesec}

\subsection{Observing Procedure}
Since we were looking for radial velocity variations among stars in clusters that might harbor exoplanetary systems, our observing procedure 
involved repeat observations of our target fields (Table~\ref{targtable1}) with M2FS. To date, we have observed our pointings in NGC~2516 and NGC~2422 12 and 10 times, respectively. We also obtained 35 observations of our principle RV standard HIP~48331 along with a small number of observations of the other comparison standards (Table~\ref{stdtargtable}). The specific dates of the observations will be summarized in a later paper describing our time-series spectroscopy; for the purposes of the present paper we simply note that the data were obtained during runs in November 2013, February 2014 and December 2014.  

As noted earlier, fibers are positioned at the focal surface of the telescope with M2FS using aluminum plug plates that are manually installed 
and plugged.  For a given field, each assigned fiber is positioned at a specific hole in the plate marked for that fiber.   We typically deployed 
128 fibers for the NGC~2516 and NGC~2422 fields, though with dead or otherwise inactive fibers excluded.   Once plugged, a typical 
observation then consists of acquiring the field using a set of ancillary fibers and imaging optics aligned to reference stars in each field.   
Typical total exposure times for each observation used in this paper were 2 hrs and 2.5 hrs per visit for NGC~2516 and NGC~2422, 
respectively. Most visits consisted of 3-5 individual exposures to aid in cosmic-ray removal and to enable measurement of the photon midpoint 
for barycentric correction. This yielded a median S/N of $\sim55$ ($\sim1\sigma$ range 15 - 70) per 1D extracted pixel 
($\sim90$ per resolution element). A detailed table of our individual observations and the targets therein will be presented in 
our next paper.  For each observation we obtain calibration data consisting of a Thorium-Argon-Neon lamp exposure and a 
quartz lamp exposure either before or following the science frame. On some nights during which our targets were observed we also 
obtain either evening or morning twilight spectra.       

RV standard observations are performed by placing a single fiber in a standard hole on the fiber plug plate and offsetting from 
the field center. Remaining fibers are left in their positions and see only sky. Typically three exposures of two minutes each are used to obtain 
a spectrum of S/N$\sim240$ per extracted 1D pixel.  Finally we obtained four epochs (1 or 2 per observing run) of the telluric 
standard calibration field in NGC~2516. These spectra have a median S/N of 160 and were obtained in 3 to 5 exposures totaling roughly one half hour.

\section{Reduction}
\label{reductionsec}
\subsection{Image Processing}
Basic data reduction follows a mostly traditional path using a custom set of Python tools written for M2FS. 
The quadrant images produced by the CCD's four amplifiers were bias corrected by subtracting the median overscan 
column and then row. We then converted counts to electrons and used
the Python implementation of the L.A. Cosmic algorithm \citep{lacosmics} to detect cosmic rays: this algorithm takes the Laplacian of the 
image and identifies cosmic rays using their steep intensity gradient. The quadrants were then packed together and stored with a
variance frame consisting of electrons plus the square of each quadrant's read noise and a bad-pixel mask. 

We created a cleaned, summed image by adding the electrons and variances of each pixel across a sequence of frames. 
Masked pixels in each component frame were repaired with their expectation value based on the other frames. 
A scaling value was computed for each frame to normalize throughput and exposure time variations by using the total 
time-normalized counts of all spectra as a proxy for throughput variability. Bad pixels in each frame were then repaired 
using the expectation value determined from the good pixels in other frames and the frame scaling
values. The variance of the final, summed pixel was inflated appropriately at every impacted pixel. 

Typically one would flat-field the resulting frames, however Quartz trace flats of comparable signal-to-noise take an 
impractical amount of observing time and M2FS does not presently have a means of obtaining uniformly illuminated 
CCD frames. Engineering work shows M2FS CCDs are free from large defects and indicate pixel-to-pixel sensitivity 
variations of about 1.7\% and only 0.25\% of pixels are significantly hot or cold. 

\begin{figure*}
\centering\includegraphics[width=1\textwidth,clip=true,trim=0cm 0cm 0cm 0cm]{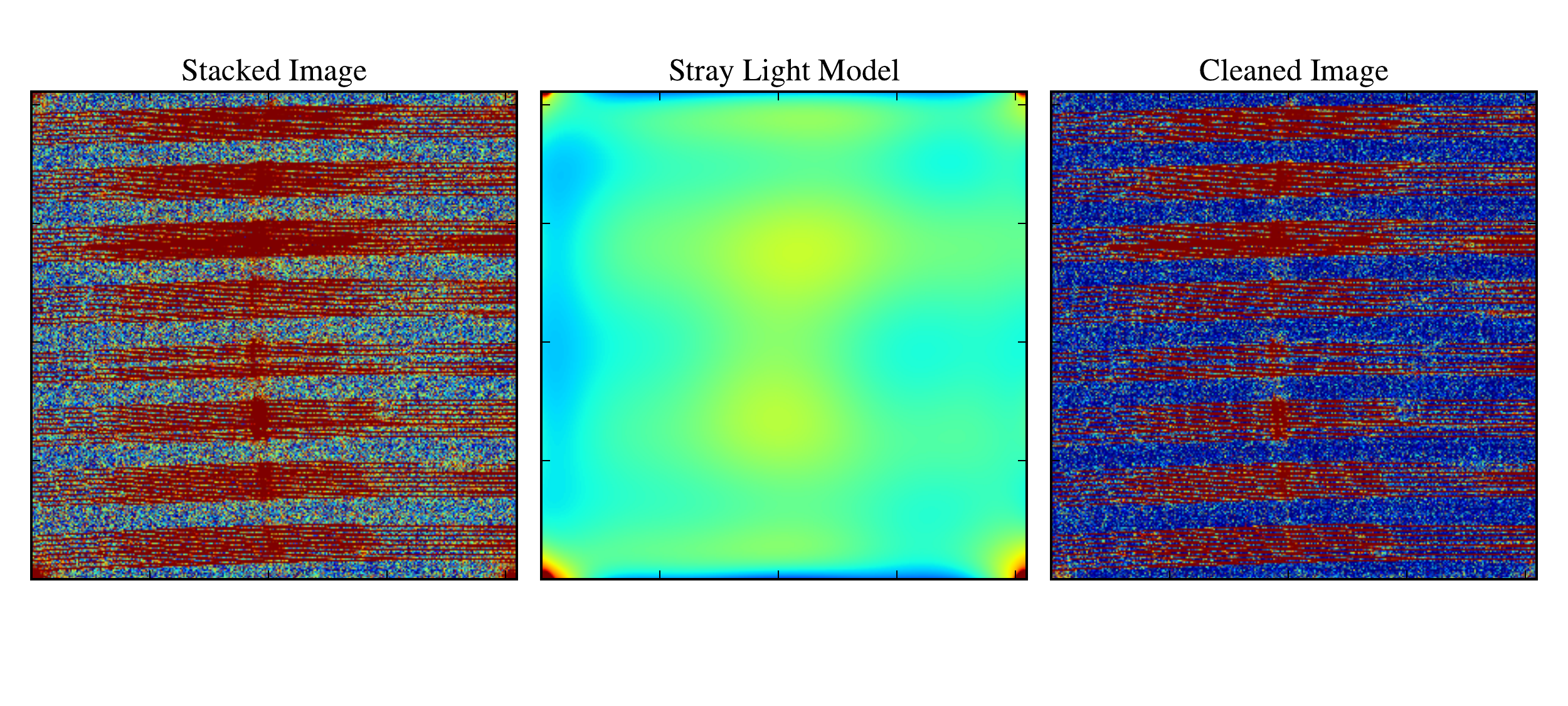}
\caption{\label{exampleim} Left to right: A stacked science frame, a map of scattered light and dark current, and the cleaned frame.
Amplifier glow and spectra are heavily clipped in the frames.  The central, bright, vertical swath stems from Littrow ghosts. All three 
images share the same color scale.}
\end{figure*}

We subtracted a combined dark current and scattered light map, an example of which is shown in Figure~\ref{exampleim}
along with the image prior to and post subtraction.
This map was computed by first modeling and removing the amplifier glow in each corner by fitting a 2D Gaussian surface. 
All remaining pixels within about a standard deviation of the mean light level in the dark regions between bundles of 16 fibers
were then selected as ``scattered light'' pixels and used to fit polynomials across the image. The resulting map was Gaussian smoothed 
using a 32 x 64 pixel rectangle ($\sim$ 1.5 x 3 fiber spacings) and subtracted from the image. Without this step, these components would 
amount to about 150 $e^-$ per 1D pixel in our extracted spectra, ranging from $\sim 5 - 50\%$ of our extracted signal for our brightest
to faintest targets.

\subsection{Extraction}
\begin{figure*}
\centering\includegraphics[width=1\textwidth,clip=true,trim=0cm 0cm 0cm 0cm]{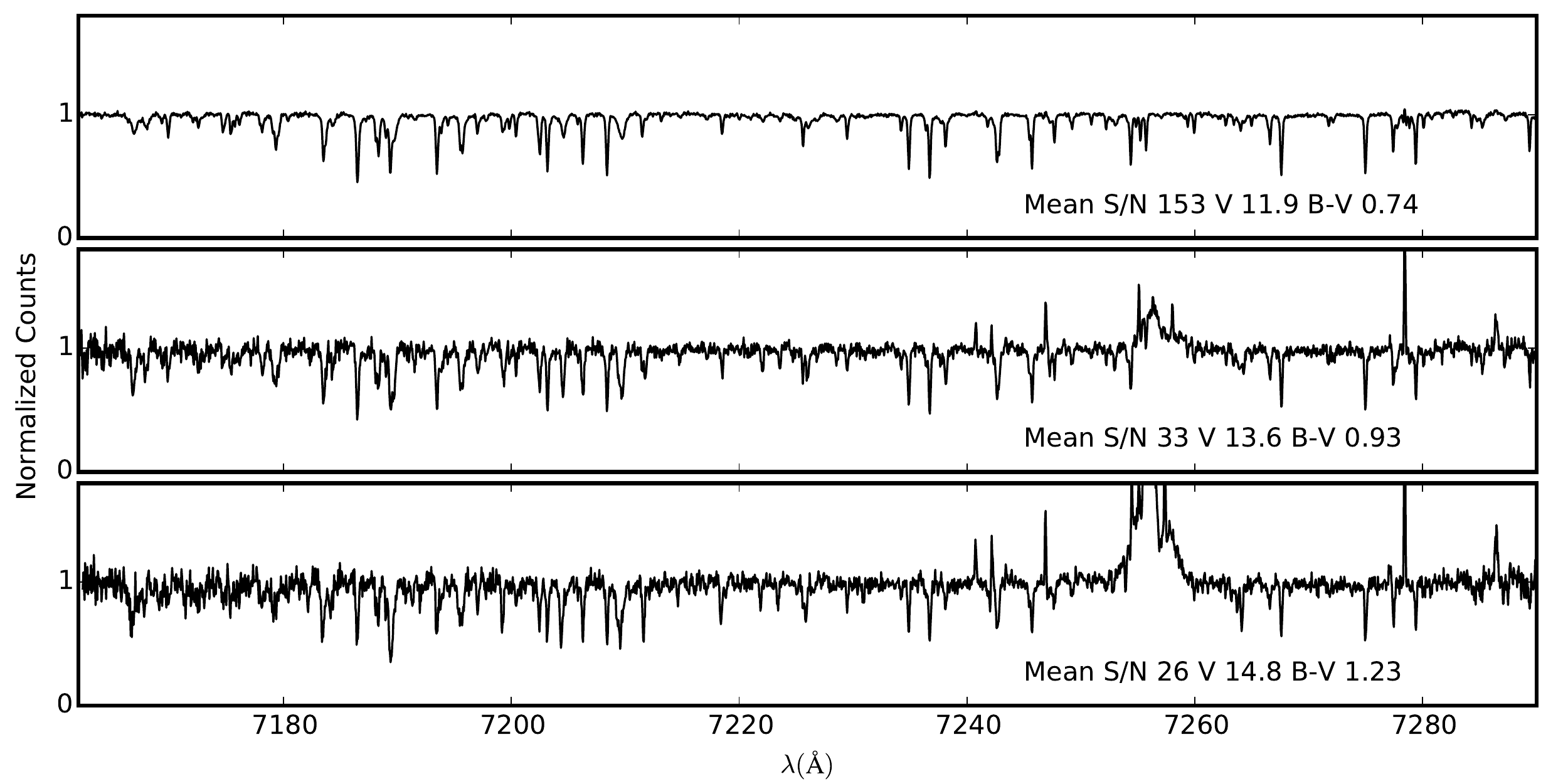}
\caption{\label{exampleextractions} Spectra of representative F5, G5, and K5 target spectra. 
Note that sky emission lines become increasingly prominent for fainter targets. 
The large defect at $\sim7255$ \AA\ is M2FS's Littrow ghost.}
\end{figure*}

Each processed frame was then extracted using the PyRAF task \texttt{apall}. We first identified the approximate aperture locations 
and traces using dome flats taken during the day with all usable fibers plugged. The apertures were then median shifted to the 
locations of the quartz traces taken with each exposure to account for any temperature drift or repositioning errors in the 
instrument. We then extracted both the science and variance frames without variance weighting using identical apertures.
Finally we continuum normalized the spectra by iteratively fitting a polynomial, each time excluding points 1 sigma below or 2 sigma above. 
An example of order 49 for a $\sim $ F5V, G5V, \&  K5V star in our sample is shown in Figure~\ref{exampleextractions}.  We do not
wavelength calibrate our spectra in a traditional sense as wavelengths are determined as part of the modeling process. 

\section{Analysis}
\label{analysis}

We measured each target's stellar properties (e.g. $T_{eff}$, [Fe/H], [$\alpha$/Fe], $v_r\sin(i)$) and line-of-sight radial velocity (RV) by 
fitting a model of each extracted, normalized spectrum to the spectrum in 1D pixel space. This approach is similar 
to the popular gas cell approach where molecular absorption lines from a well-calibrated gas cell (e.g. $\rr{I}_2$, Ammonia) are used
as a simultaneous probe of pixel wavelengths and the instrumental point-spread function (PSF). Here we make use of the abundant 
atmospheric $\rr{H_2O}$ lines in the 7230 \AA\ region as the imprint of a giant gas ``cell.'' A variant of this idea was originally 
proposed by \citet{griffin} more than 40 years ago and has been used with success to measure RVs 
in both the optical \citep{cochran88,2010A&A...515A.106F} and the infrared 
\citep{2007ApJ...666.1198B,2008ApJ...687L.103P,seifahrt10,2010ApJ...723..684B,2011ApJ...735...78C,ysasref}. These studies 
have demonstrated that telluric 
features are stable to 10 m/s. This should not come as a surprise as all of the water and the general bulk of our atmosphere is within 
the first 8 - 16 km where typical bulk motions are below 10 m/s and not along the line-of-sight. We will quantify this source of uncertainty  
when we discuss our achieved RV precision in Section~\ref{rvprecisionsec}.
 
The model is
constructed by combining a template of the telluric absorption spectrum, $T(\lambda)$, one or more synthetic stellar spectra, 
$S(\lambda;T_{eff},\rr{[Fe/H]},\rr{[\alpha/Fe]}, v_r\sin(i), \rr{RV})$, a synthetic sky emission spectrum, $Sky(\lambda)$, and a Solar spectrum 
$Sun(\lambda;  RV_{\sun})$.
\begin{multline}
\label{modeleq1}
M(\lambda)=T(\lambda)^\alpha\cdot(\\S(\lambda;T_{eff},\rr{[Fe/H]},[\alpha/\rr{Fe}], v_r\sin(i), \rr{RV})^\beta+\\\gamma\cdot Sun(\lambda+s;RV_{\sun}))+\eta\cdot Sky(\lambda) 
\end{multline}
This model is then resampled onto pixels, convolved with a model of the 1D projection of the PSF, and normalized.
\begin{equation}
\label{modeleq2}
M(pixel)=\frac{PSF(pixel;\boldsymbol{\sigma})*M(pixel)}{N(pixel;\boldsymbol{\zeta})}
\end{equation}
In the above equations the scalars $\alpha,\ \beta,\ \gamma,\ \eta,\ s,$ and vectors $\boldsymbol{\sigma}$ and $\boldsymbol{\zeta}$ are 
 model parameters which will be described in the next section and are summarized in Table~\ref{modeltabletable}.
 
\begin{deluxetable}{ccc l}
\tablewidth{0pc}
\tablecaption{Model Parameters\label{modeltabletable}}
\tablehead
{

\colhead{Component} & 
\colhead{Symbol} & 
\colhead{Number} &
\colhead{Comments}}
\startdata

Wavelength &  $\boldsymbol{\mu}$ & $\leq$8 & $\lambda(pixel)=\sum_i^8\mu_i L_i(pixel)$   \\
PSF &  $\boldsymbol{\sigma}$ & 1& The PSF FWHM\\
&& 3 & $FWHM(pixel)=\sum_i^2\sigma_iL_i(pixel)$ \\
&& 5 &  A 5th order Hermite parameterization\\
&& 22 & \citet{attaining3ms} parameterization  \\
Nomalization & $\boldsymbol{\zeta}$ & 12 & $norm(pixel)=\sum_i^{11}\zeta_i L_i(pixel)$ \\

Stellar Temperature & $T_{eff}$&  1 & Snaps to 100~K grid\tablenotemark{a} \\ 
Iron Abundance & [Fe/H] &  1 & Snaps to 0.1 dex grid\\ 
Alpha Abundance & [$\alpha$/Fe] &  1 & Snaps to 0.1 dex grid \\ 
Stellar Rotation & $v_r\sin(i)$ & 1 &  \\
Radial Velocity & RV & 1 & \\

Airmass & $\alpha$ & 1 & Scale atmospheric transmission \\
Veiling & $\beta$ & 1& Scales stellar absorption lines in unison.\\
& & & Held at unity when fitting stellar parameters\\
Solar Flux & $\gamma$ & 1& Fractional contribution of solar flux\\
Solar RV & $ \rr{RV}_{\sun}$ & 1 & \\
Solar Offset & s & 1 &  Offset between PHOENIX and Kurucz wavelengths\\ 
Sky Emission & $\eta$ & 1 & Scale SkyCalc spectrum\\

\enddata
\tablenotetext{a}{Template spectrum with nearest value is used.}
\end{deluxetable}

\subsection{Model Input}
\subsubsection{Stellar Light}
\label{phoenixgridsec}
Our pipeline uses the PHOENIX grid as it samples a large stellar parameter space (far beyond our region of immediate interest), 
is the successor to the grid used in \citet{ysasref}, and in no small part because it is the grid for which we attained 
the best RV precision for our RV standard. 
To verify this point we also checked both the \citet{Coelho:2014cd} grid and the AMBRE grid 
\citep{ambre} and found both to result in larger RV measurement errors for our RV standard. Prior to 
modeling we up-sampled the PHOENIX grid (see Table~\ref{gridtable}) in the parameter space relevant to our target stars using 
linear interpolation by way of the SciPy function \texttt{map\_coordinates}. Library spectra are normed by the maximum 
continuum value in the fitting region and linearly interpolated onto a constant $d\log(\lambda)$ grid, 
adopting the largest step size present in the raw spectrum, just prior to use in the modeling pipeline.

During fitting, the surface gravity is tied to the effective temperature via Eqn. \ref{tempmasslumeqn} which is derived from the 
mass-luminosity, temperature-luminosity, and mass-radius relations for lower main-sequence stars as our fits do 
not appear to be particularly sensitive to variations in $\log(g)$ among our main sequence targets. This weak dependence is 
seen in other techniques as well:  \citet{Casagrande11} reports that even variations as large as a half-dex affect 
$T_{eff}$ by only a few tens of Kelvin. 
We calibrated the relation using Solar values corrected for age per the plot in \citet{ribas10}, $L=0.85 L_{\sun}$, $R=0.925R_{\sun}$.
\notetoeditor{This equation works in Manuscript form but not in two column mode. I'm not sure how to fix it. 
I'd suggest it either spanning the columns or I'll have to figure out how to break it up.
}
\begin{eqnarray} 
R\propto M^{0.9} \\
L=L_{\sun} (\frac{M}{M_{\sun}})^{4} \\
L=4\pi R^{2} \sigma T_{eff}^{4}\\
g=\left(\frac{L_{\sun}}{4 \pi  \sigma }\right)^{4/11}\frac{ G M_{\sun}}{ R_{\sun}^{30/11} T_{eff}^{16/11}}\\
\log(g)=\log(\frac{9.44\times10^{9}}{T_{eff}^{16/11}})\label{tempmasslumeqn}
\end{eqnarray}
This input brings with it the astrophysical parameters $T_{eff}$, [Fe/H], [$\alpha$/Fe], $v_r\sin(i)$, and RV along with a feature depth parameter $\beta$ 
which allows fudging the optical depths of all the stellar lines in unison (we note that this is a simplification as lines are not expected to scale in unison). In the event of a spectroscopic binary we can enable 
multi-component modeling, using two sets of these parameters and an additional multiplicative parameter for the ratio of
flux received from the two stars.

\begin{deluxetable}{ccccc}
\tablewidth{0pc}
\tablecaption{Synthetic Grid Spacing\label{gridtable}}
\tablehead
{

\colhead{Grid} & 
\colhead{$\Delta T_{eff} (K)$} & 
\colhead{$\Delta \log(g)  (dex)$} &
\colhead{$\Delta [Fe/H] (dex)$} &
 \colhead{$\Delta [\alpha/Fe] (dex)$}}
\startdata

PHOENIX &  100 & 0.5 & 0.5 & 0.2   \\
Resampled PHOENIX &  100 & 0.1 & 0.1 & 0.1\\
\enddata
\end{deluxetable}

\subsubsection{Telluric Transmission}
We considered two options for the telluric transmission model: the NSO empiric transmission spectrum \citep{wallaceref} and the 
synthetic TAPAS model \citep{tapasref}, nominally tailor made for the atmospheric conditions during each observation. The NSO spectrum derives
from data obtained on the McMath-Pierce solar telescope using the Fourier transform spectrograph (FTS) in the late 1980s. 
TAPAS spectra are computed as described in \citet{tapasref} for the conditions of each exposure. 

When using TAPAS spectra in constructing our models for HIP 48331 we measured an RV $286\pm8$~m/s larger than when using the 
NSO spectrum as the template and also observed a reduction in RV precision (c.f. Section~\ref{rvprecisionsec}). 
This shift is on the order of the uncertainty of the $\rr{O_2}$ and $\rr{H_2O}$ line positions in the HiTran database 
\citep{hitran} and so is perhaps not unexpected given that is is used as the data source for TAPAS. 
It is interesting to note that this is also of order the shift caused by mixing \citet{edlen} and \citet{ciddorref} 
air/vacuum relations (e.g. converting one way with Ciddor and back with Edl\'{e}n), however we are unable to 
ascertain a list of conversions applied between the original data and the output TAPAS spectrum and are
unable to offer any firm conclusions regarding the source of the shift. We do 
measure better $\chi^2$ values when using TAPAS spectra and suggest that the TAPAS pipeline models
differences in atmospheric line strengths between Kitt Peak and Las Campanas well. Given the reduced RV 
precision we used the NSO FTS data for our analysis. This input brings with it the parameter $\alpha$ to 
logarithmically scale the absorption features as a proxy for airmass. 

\subsubsection{Sky Emission}
We used the ESO SkyCalc tool \citep{skycalc1,skycalc2} to obtain night sky emission spectra for our wavelength region. These spectra match the locations of 
the night sky emission lines well, though they do not always perfectly match their relative strengths. This input adds a multiplicative 
scaling parameter, $\eta$, to adjust the predicted count rate.

\subsubsection{Instrumental Effects}
The modeling code also includes inputs for the instrumental dispersion relation, point spread function (PSF), and allows for inaccuracies in our 
continuum normalization. The dispersion relation is a set of Legendre polynomial coefficients ($\boldsymbol{\mu}$) which 
yield the wavelengths at each extracted pixel. We used a second set of Legendre polynomial coefficients ($\boldsymbol{\zeta}$) 
to compute a normalization polynomial that accommodates errors in continuum normalization during extraction.

The PSF is widely understood \citep[c.f. e.g.][]{attaining3ms, beancrires, ysasref} to have a significant impact on the precision with which line 
centroids can be recovered and ultimately the RV precision. We investigated this effect by modeling our PSF with a 
Gaussian, a Gaussian with width quadratically varying along the order, the multi-Gaussian parameterization of \citet{attaining3ms}, 
and a 5th order Gauss-Hermite series \citep{gao} and then determining which prescription minimized the RV standard deviation for our RV standard. 

\subsection{Modeling Process}
\subsubsection{Model Construction}
\label{modelingprocesssec}
To compute the model described in Equations \ref{modeleq1} and \ref{modeleq2} the code will first fetch the synthetic spectrum of 
nearest temperature, surface gravity, iron, and $\alpha$-element abundance from our grid 
(recall $\log(g)$ is computed per Eqn. \ref{tempmasslumeqn}) along with the telluric absorption, Solar, and pointing dependent emission
spectra. These spectra are all in excess of $R\sim 600,000$. 

The telluric transmission spectrum is scaled logarithmically by $\alpha$ and the sky emission and Solar spectra are scaled 
multiplicatively by $\eta$ and $\gamma$, respectively. The solar spectrum is Doppler shifted
by multiplying the wavelength grid by the appropriate Doppler factor.

The stellar spectrum is normalized by the maximum flux in the wavelength region to perform a simple continuum normalization 
while preserving the slight blackbody effect in our narrow region. In the case of a multi-star fit normalization of each spectrum is still carried out in this manner. The spectrum is scaled logarithmically by $\beta$ to account for any mean discrepancy in line depth and is then rotationally broadened via convolution with a kernel computed based on Equation 17.12 in \citet{greyphotospheres}\footnotemark[1]. Finally the spectrum is Doppler shifted by multiplying the wavelength grid by the Doppler factor. 

\footnotetext[1]{We do not use the
\texttt{lsf\_rotate} library function from \citet{hubenylanz} as there are numerical and functional errors in the construction 
of the kernel that cause discontinuities in $\chi^2$ as a function of $v_r\sin(i)$.}

When computing a binary model this process is
repeated for each star and a weighted average of the two spectra is taken after they are placed on the sub pixel grid in the following step.
In this situation the weight is an additional free parameter (constrained between 0 and 1). We investigated combining the stellar components in a 
manner that enforces their relative flux from the synthetic library however this often led to grossly inappropriate minima. When modeling a telluric 
standard spectrum we model the stellar component as unity. 

If including scattered Solar light the solar spectrum is multiplicatively scaled by $\gamma$, s is added to the Solar wavelength grid, and then
the grid is multiplied by the appropriate Doppler factor. 

Each of the components is linearly interpolated onto a $10^{th}$ pixel wavelength grid computed from the dispersion relation. 
We note that cubic spline interpolation significantly decreased our RV stability. These component spectra are then combined as in 
Eqn. \ref{modeleq1}. 

This model is then convolved with the PSF kernel representing the instrument's point spread function as
mangled by our simplistic extraction. The kernel size was selected such that the kernel is less than $10^{-4}$ at the window 
edge and the kernel is always constrained to be positive. To maintain a modicum of speed pixel dependent convolutions are carried 
out via a FORTRAN subprogram.  When using an asymmetric PSF we noted that the center of the enclosed power was not 
generally located at the central sub-pixel nor at some constant offset. We observed a shift for typical, good fits of $1.5\pm0.86$ pixels redward with
the Hermite parameterization. 
Constraining the centroid to $\pm0.05$ pixels of the center confers substantial improvements to RV precision (See Section~\ref{rvprecisionsec}).
In the limit of a linear dispersion relation this effect would correspond to a simple pixel shift and would not be expected to have 
any impact on results, however our wavelength solutions are not linear. Our code constrains the centroid by shifting the PSF kernel and 
hence constrain the center of enclosed power to within $\sim100$ m/s of the central sub-pixel. Interpolated shifts or a PSF parameterization 
which constrains the enclosed power may be worth future investigation. 

After convolution, the sub-pixel values are averaged to yield pixel values and the model is divided by the normalization polynomial, yielding
a model spectrum.

\subsubsection{Merit Function}
\begin{figure}
\plotone{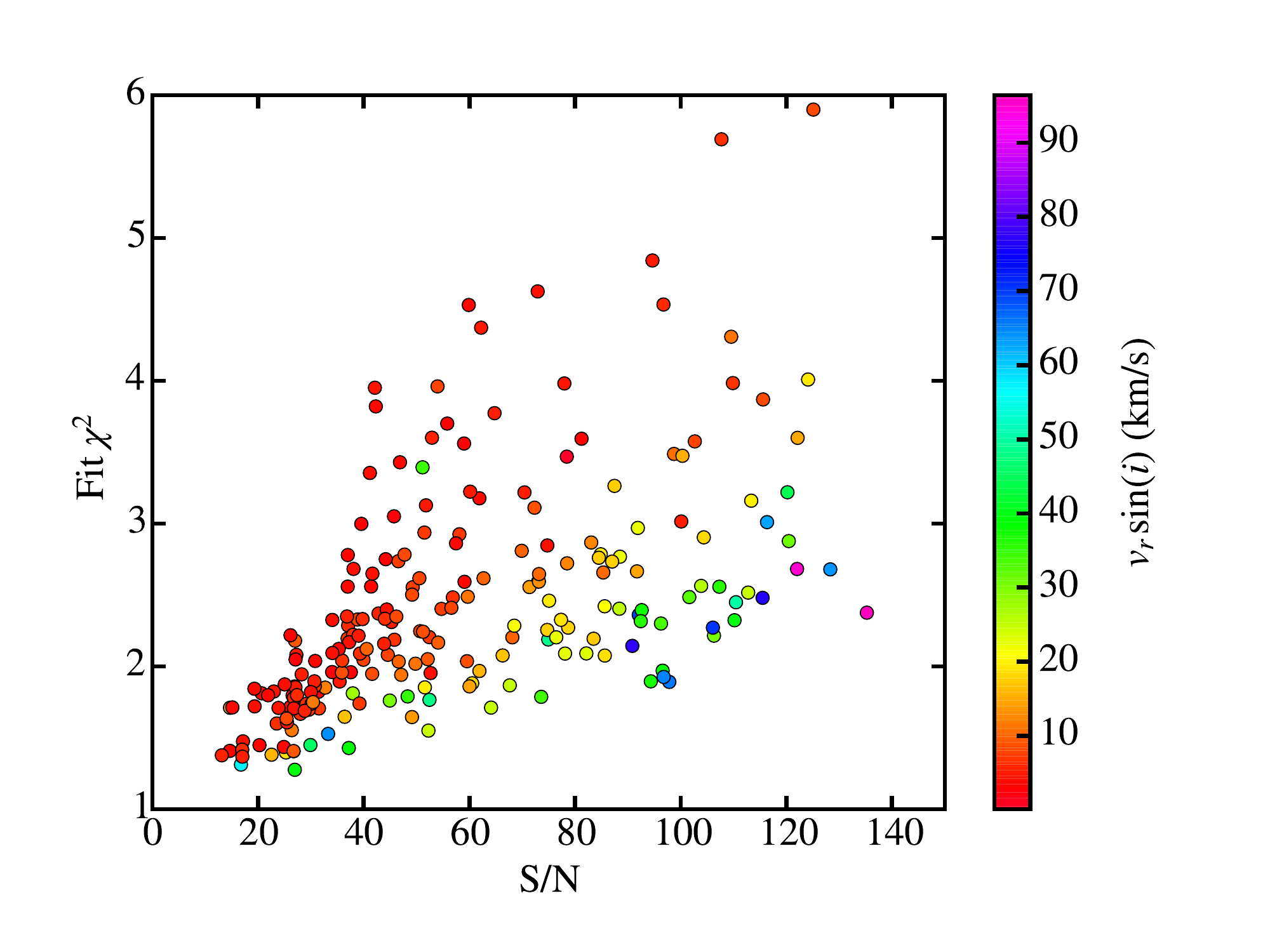}
\caption{\label{cstrend}
This figure shows the average best-fit reduced $\chi^2$ for each of our cluster stars plotted as a function of S/N.
Colors denote the rotational velocity measured for each star. There is a distinct upward trend in $\chi^2$ with S/N.}
\end{figure}

To determine the optimal model, we used the $\chi^2$ as a merit function. The fitter computes the weighted 
mean square error of our model with the normalized spectrum, masking pixels based on a wavelength mask (e.g. for sky lines, if desired), 
an RV dependent wavelength mask for stellar lines, and a pixel mask for detector defects and the Littrow ghosts. Wavelength masks are additive:
for stability, once masked, a change to the wavelength solution will not cause a pixel to unmask. 
The weights are computed as the ratio of the square of the continuum normalization to the variance spectrum. 
We noted a significant upward trend in our best-fit $\chi^2$ with increasing signal (c.f. Figure~\ref{cstrend}), which 
we attribute to an improved ability to identify finer errors in our computed model.  Visual inspection shows that PHOENIX spectra 
consistently mismatch stellar line depths and with increasing S/N these mismatches become increasingly significant. 

\subsubsection{Model Fitting}
\label{modelfittingsec}
Prior to fitting we visually reviewed all $\sim2700$ spectra and excluded spectra with average continuum S/N less than 12 (per pixel). 
We limited all our fits to the extent of order 49 with continuum S/N greater than 12 or the columns between pixels 25 and 2700 
($\sim7160 - 7290$ \AA). These column extents were selected such that we have a slight margin at either end with which to estimate 
the continuum level at the order edges beyond the fitting extent.  Generally fits were to all 2675 pixels in this region. We masked pixels affected by 
extraction artifacts, continuum normalization errors, uncorrected cosmic rays, or Littrow ghosts. After fitting we inspected the results
for any failures (generally due to a poor initial RV guess) and either corrected them, flagged them to handle as exceptional cases, or 
excluded them from analysis.

We optimized the model in stages. 1) For a subset of spectra we first obtained an initial guess for the wavelength 
solution by eye. These guesses were used to bootstrap initial relations for the wavelength solution in one frame of each arm on each run. 
The initial PSF width was chosen such that it coincides with M2FS's nominal resolving power in our configuration without asymmetry and
an initial spectral type was chosen assuming Solar abundance and using the (B-V)-$T_{eff}$ relation of \citet{teffcolor} with 
reddening corrected values for B-V. These parameters were then used as the initial 
values for a round of fits from which we constructed a predictive model of the dispersion parameters as function of M2FS arm, CCD 
trace position, and night. We found the $4^{th}-7^{th}$ order wavelength parameters are neither a function of instrument temperature, 
(mis)focus, or run and thus adopted a simple polynomial model as a function of aperture position based on the best fit values
for all $\sim2600$ usable spectra. We only adopt the mean values as an initial guess for the $1^{st}-3^{rd}$ order parameters and 
note that the wavelength zero point parameter is predicted for each exposure separately. 
2) We then refit all our spectra with the initial wavelength parameters determined in phase one, holding the 
$4^{th}-7^{th}$ order wavelength parameters fixed, and adopted the inverse variance weighted means of the best-fit spectral type 
parameters and $v_r\sin(i)$ values for each star. 
3) We performed a final round of fits still holding higher order wavelength parameters fixed, now along with the spectral type 
parameters and $v_r\sin(i)$. We used the RVs from this final fit as the values we adopt for each star. Except in the case of 
large amplitude binaries the RV was always started from the adopted multi-epoch mean.

\begin{figure*}
\centering\includegraphics[width=.9\textwidth,clip=true,trim=0cm 0cm 0cm 0cm]{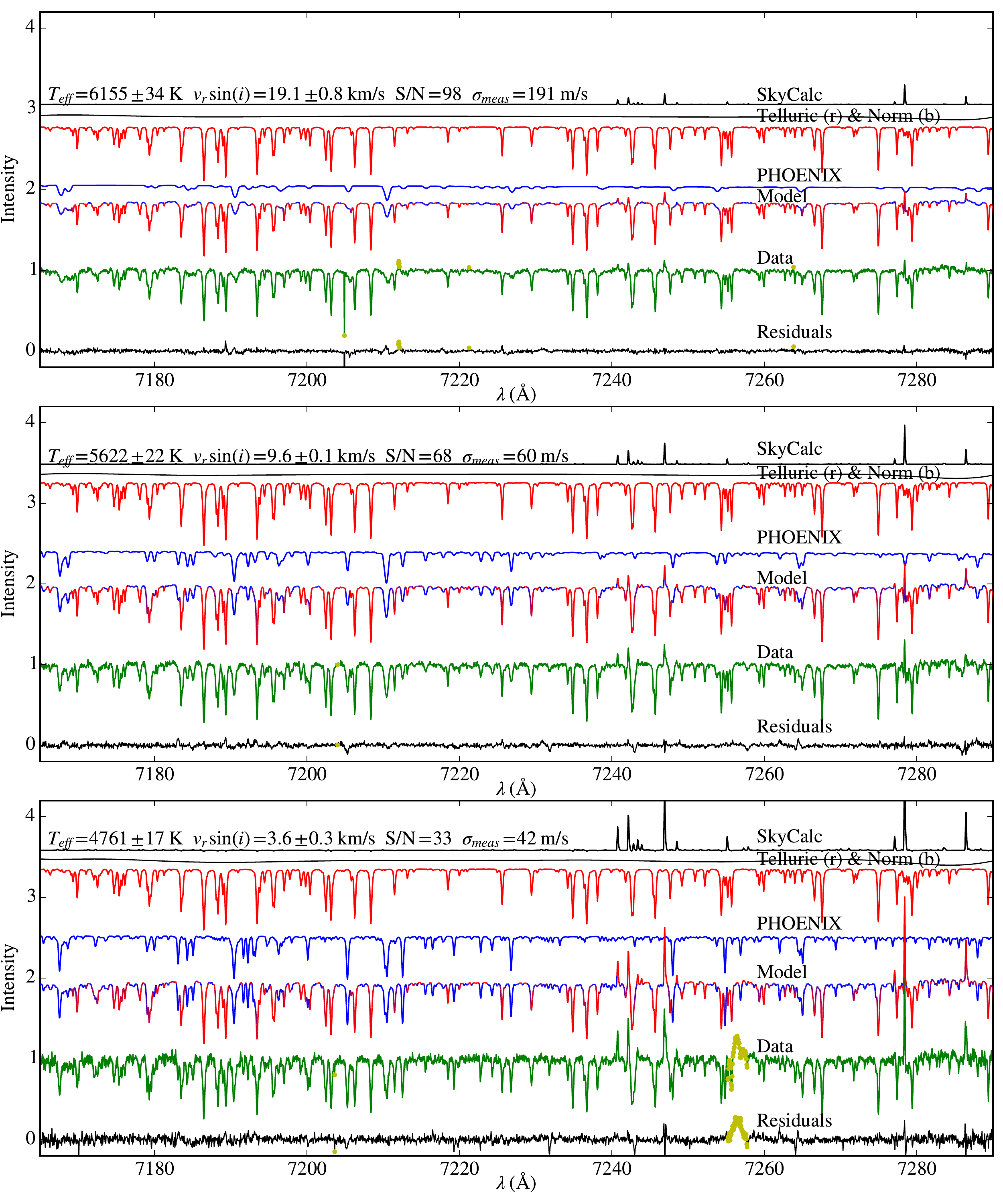}
\caption{\label{fitexamplefig} Examples of fits to F (top), G (middle), and K (bottom) stars in our sample. Each spectrum is labeled with the adopted $T_{eff}$, and $v_r\sin(i)$ for the star along with mean S/N and $\sigma_{meas}$ for the plotted spectrum. From top to bottom in a single plot we show the sky emission component (black), the normalization curve (black), the FTS telluric transmission profile (red), The PHOENIX model (blue), our resulting model of the data (blue and red), the extracted spectrum (green), and the residuals (black). Data and residuals that are masked have yellow points overlaid.}
\end{figure*}

Section~\ref{resultssec} describes the precision with which we recover these values.  As our initial wavelength guesses are crude we 
investigated the impact priming our wavelength solution 
parameters with ThAr calibration fits to verify we were not introducing a fitting bias. For this test we refit all of 
the RV standard spectra using the IRAF identify task solutions to corresponding ThAr data for the initial wavelength 
solution. We found no impact. Holding the
higher order wavelength parameters fixed improves RV precision by approximately $\sim5$ m/s at all signal-to-noise 
levels (c.f. Sec. \ref{rvprecisionsec}). An example fit is shown in Figure~\ref{fitexamplefig}.

\epsscale{1}

Optimization was carried out using the \texttt{MPFIT} \citep{mpfitref} package to minimize the weighted errors for each unmasked 
pixel. In previous iterations of our software we used the AMOEBA minimizer: the downhill-simplex optimizer \citep{neldermead65} appears more 
tolerant of poor initial guesses, but minimization takes a greater number of function evaluations, does not yield a parameter covariance matrix, 
and requires parameter limits be hacked on as they are not inherent to the algorithm. We also investigated using the MCMC core of 
\citet{exofast}, and briefly explored a genetic fitter, the latter of which proved to be of similar quality but highly inefficient. We found 
these various methods of optimization to all be of comparable end result but with significantly prolonged computation time.

\section{Results and Performance}
\label{resultssec}

\subsection{Stellar Properties}

\begin{figure}
\plotone{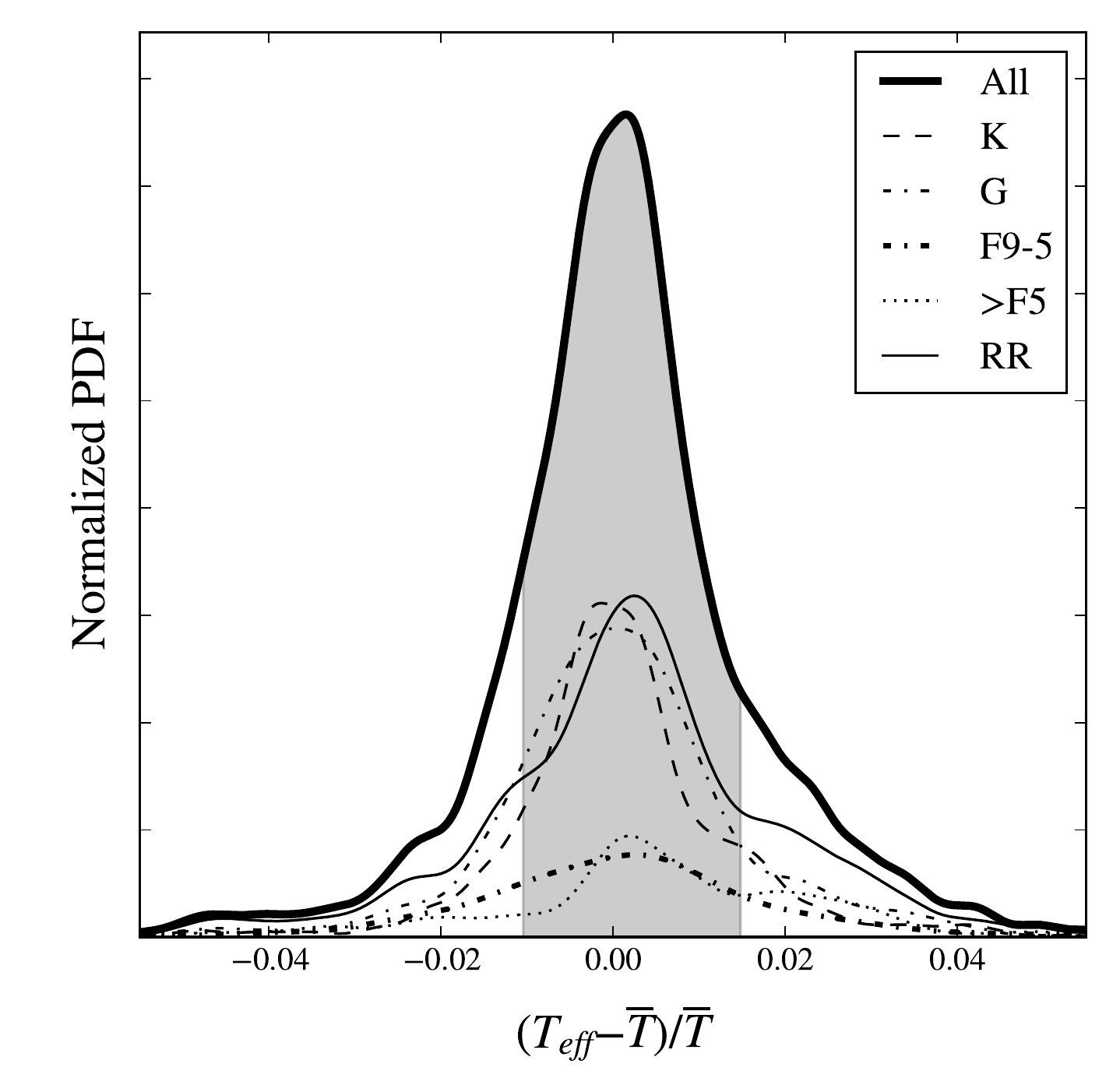}
\caption{\label{teffprecision}
This plot shows normalized PDFs for fractional $\Delta T_{eff}$ for the entire sample and subsamples of stars as a function of spectral type and 
stars with $v_r\sin(i)>8$ km/s (RR). The shaded region corresponds to $1\sigma$ for the entire sample.
}
\end{figure}

\begin{figure}
\plotone{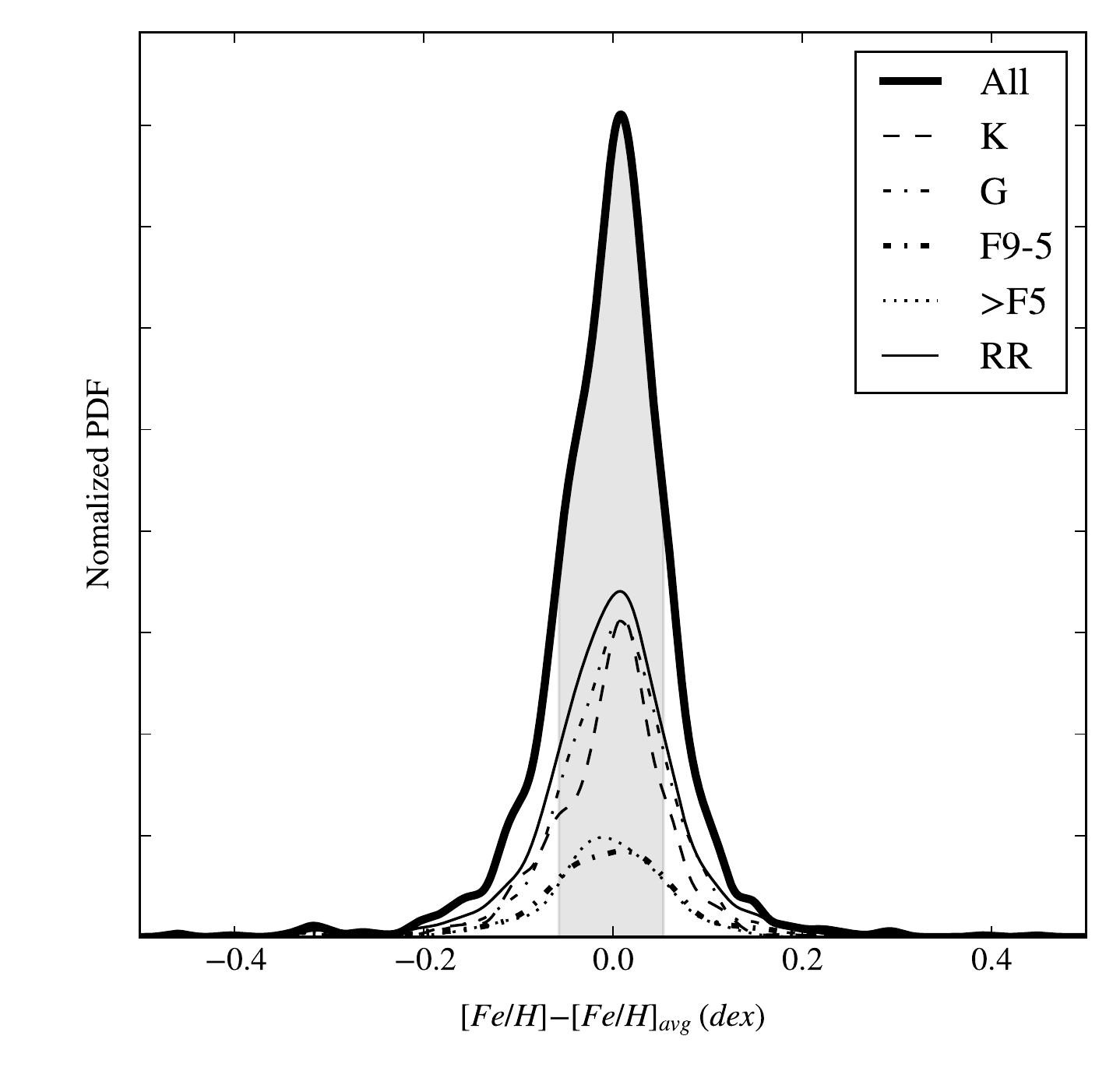} 
\caption{\label{fehprecision}
This plot shows normalized PDFs for $\Delta$[Fe/H] for the entire sample as well as subsamples based on spectral type and 
stars with $v_r\sin(i)>8$ km/s (RR). The shaded region corresponds to $1\sigma$ for the entire sample.
}
\end{figure}

\begin{figure}
\plotone{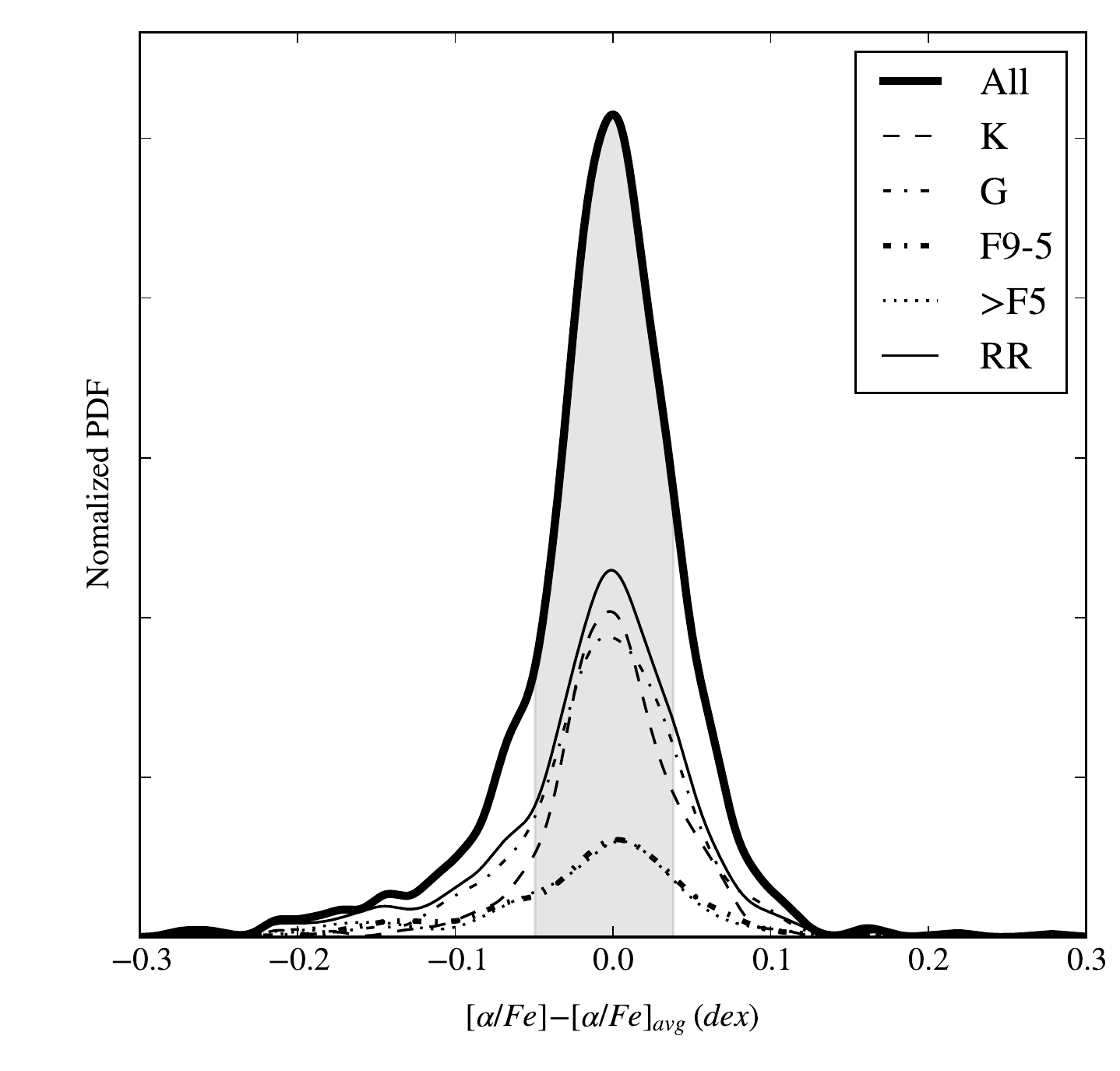}
\caption{\label{alphaprecision}
This plot shows normalized PDFs for $\Delta$[$\alpha$/Fe] for the entire sample as well as subsamples based on spectral type and 
stars with $v_r\sin(i)>8$ km/s (RR). The shaded region corresponds to $1\sigma$ for the entire sample.
}
\end{figure}

\begin{figure}
\plotone{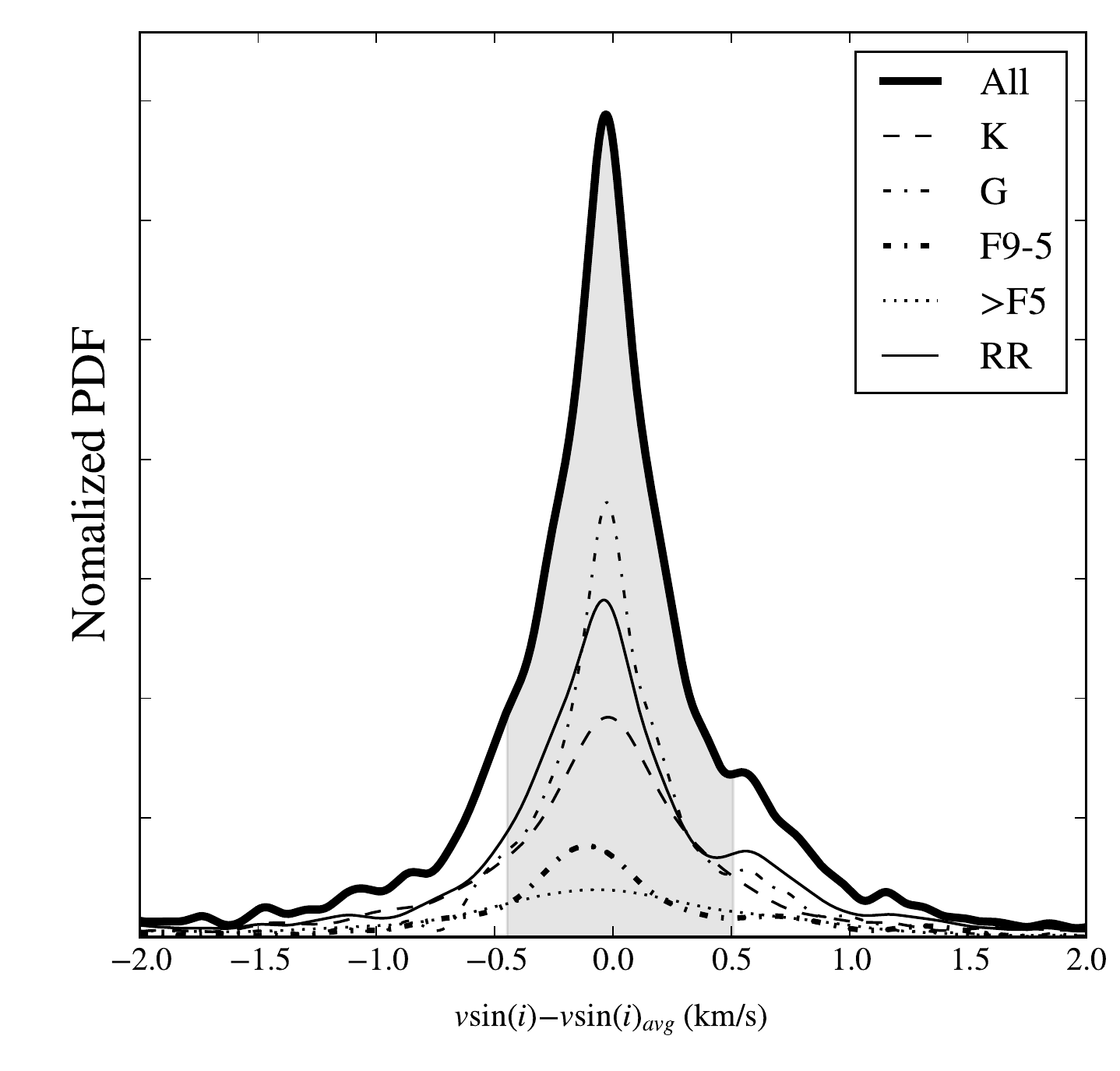}
\caption{\label{vsiniprecision}
This plot shows normalized PDFs for $\Delta v_r\sin(i)$ for the entire sample as well as subsamples based on spectral type and 
stars with $v_r\sin(i)>8$ km/s (RR)..
}
\end{figure}

We estimated the statistical uncertainty of our stellar parameters from the distribution of best fit values relative to their multi-epoch means. We used
the results from the second stage of our fitting pipeline (where stellar properties are allowed to vary from epoch to epoch, 
c.f Sec. \ref{modelfittingsec}), exclusive of spectroscopic binaries, to obtain fits to 2283 spectra of 214 targets 
(see Table~\ref{prectargtable}) with which we computed the differences between the single epoch values and the 
adopted multi-epoch values of the stellar parameter. We then performed kernel density estimation on each parameter and computed confidence intervals. 
The resulting PDFs are shown in Figures \ref{teffprecision}, \ref{fehprecision}, \ref{alphaprecision}, 
and \ref{vsiniprecision} and the $1\sigma$ confidence intervals given in Table~\ref{stellarpropprectable}. In addition to the overall PDFs,
the plots give PDFs for subsamples grouped by spectral type and rotation rate ($v_r\sin(i)>8$ km/s). The selection of 8~km/s as a grouping is largely 
arbitrary, though corresponds with the point where $v_r\sin(i)$ becomes the dominate source of line broadening relative to the instrumental PSF.
Though we do not explicitly show PDFs for groupings in S/N, the PDFs for spectral type show this by proxy; later spectral types are fainter and have lower S/N spectra. We saw no evidence that the PSF form (e.g. Gaussian vs. Hermite series) affected the measured values or their uncertainties. 

We estimated the accuracy of our technique for determining $T_{eff}$, [Fe/H], and [$\alpha$/Fe] by comparing the values we measured with those in the literature. 
Tables \ref{stdparvals} gives our values for the six standard stars along with those from the literature. 
Tables \ref{stderrors} then reports the differences in these values: we adopt the averages therein as an estimate of our systematic 
uncertainties. As an additional check we fitted $\sim900$ twilight spectra and report the values and differences thereby obtained. 
Though these values are of comparable quality, we excluded them from our average as the large number of spectra would 
heavily bias the results. 
We tested our $v_r\sin(i)$ accuracy by comparing our values with those reported in \citet{n2516activity}, with which we have thirty-seven
targets in NGC~2516 in common\footnotemark[2]. Our values agree to within 5 km/s for all but four stars, which are all 
spectroscopic binaries. For the remaining 33 stars our adopted, multi-epoch mean values agree with a standard deviation of 2.2 km/s. 
We recovered the correct $v_r\sin(i)$ to better than 0.1 km/s in fits to our twilight spectra. 
Our code is not able to reliably measure $v_r\sin(i)$ values below $\sim2$ km/s (roughly one third of our velocity resolution).

\footnotetext[2]{We do not perform this exercise on $T_{eff}$ as the values reported are from colors.}

Based on this analysis, we report our fits yield typical single-epoch precisions of 75 K, 0.05 dex, and 0.75 km/s for 
$T_{eff}$, [Fe/H] and [$\alpha$/Fe], and $v_r\sin(i)$. We find this translates to mean multi-epoch precisions of $\pm30$~K, 
$\pm0.02$~dex for both [Fe/H] and [$\alpha$/Fe], and $\pm$0.3~km/s for $v_r\sin(i)$.
Our $T_{eff}$ values are typically cooler than available literature data for our standards by $\sim$25 K and we find a similar offset when 
fitting twilight spectra. Iron abundance values appear elevated by a tenth dex but are driven entirely by HIP 48331: 
excluding HIP 48331 $\Delta$[Fe/H] becomes $-0.03\pm0.03$ dex, consistent with our twilight fits. 
We do not see any evidence of a systematic offset in [$\alpha$/Fe] or $v_r\sin(i)$.

\begin{deluxetable}{lccccc}
\tablewidth{0pc}
\tablecaption{Precision Targets\label{prectargtable}}
\tablehead{
\colhead{Cluster} &
  
\colhead{V} &
\colhead{B-V} &
\colhead{Used} &
\colhead{Excluded} 
}
\startdata
NGC~2516  & 11.68 -- 15.09 & 0.46 -- 1.17 & 108 &18 \\
NGC~2422  & 12.19 -- 16.05 & 0.45 -- 1.31  & 106 & 19 \\ 
Total  & 11.68 -- 16.05 & 0.45 -- 1.31 &  214 & 37 \\
\enddata
\tablecomments{This table gives the ranges in magnitudes,  colors, and number of cluster stars 
used to determine our statistical uncertainties in $T_{eff}$, [Fe/H], [$\alpha$/Fe], and $v_r\sin(i)$.
We exclude spectroscopic binaries in this analysis.}
\end{deluxetable}
 
\begin{deluxetable}{lcccccccc}
\tablewidth{0pc}
\tablecaption{Single Epoch Property Precision \label{stellarpropprectable}}
\tablehead
{

\colhead{} & 
\multicolumn{2}{c}{K5-G7} & 
\multicolumn{2}{c}{G7-G2} &
\multicolumn{2}{c}{G1-F5} & 
\multicolumn{2}{c}{F4 and hotter}
\\
\colhead{Property} & 
\colhead{Lower} & 
\colhead{Upper} &
\colhead{Lower} & 
\colhead{Upper} &
\colhead{Lower} & 
\colhead{Upper} &
\colhead{Lower} & 
\colhead{Upper}
}
\startdata
$T_{eff}$ (K) &  $-$44 & +59 & $-$62 & +74 & $-$95 & +121 & $-$82 & +163   \\
$[Fe/H]$ (dex) & $-$0.06 & +0.05 & $-$0.05 & +0.06 & $-$0.07 & +0.06 & $-$0.07 & +0.05 \\
$[\alpha/Fe]$ (dex) &  $-$0.04 & +0.04 & $-$0.05 & +0.04 & $-$0.07 & +0.04 & $-$0.07 & +0.04 \\
$v_r\sin(i)$ (km/s)&  $-$0.5 & +0.5 & $-$0.3 & +0.4 & $-$0.5 & +0.7 &  $-$1.0 & +0.9\\
\cline{1-9}\\
Mean S/N & \multicolumn{2}{c}{30} & \multicolumn{2}{c}{60} & \multicolumn{2}{c}{80} & \multicolumn{2}{c}{100} 
\enddata
\tablecomments{
This table gives the upper and lower limits enclosing the central 66\% confidence interval for 
$T_{eff}$, [Fe/H], [$\alpha$/Fe], and $v_r\sin(i)$. That is, given measurement of a single epoch, there is a 66\% 
chance we would measure a value within the stated limits for a second epoch.
Below each pair of columns we list the mean S/N of stars in each group. 
}
\end{deluxetable}

\begin{deluxetable}{lrrrrr}
\tablewidth{0pc}
\tabletypesize{\scriptsize}
\tablecaption{Measured Standard Star Properties \label{stdparvals}}
\tablehead
{&
\colhead{$T_{eff}$} &
&
\colhead{[Fe/H]} &
\colhead{$[\alpha/Fe]$} 
\\
\colhead{Ref} &
\colhead{(K)} &
\colhead{log(g)} &
\colhead{(dex)} &
\colhead{(dex)} 
}
\startdata

\multicolumn{5}{c}{\textbf{HIP 48331 N=35 K5V}}\\
This work & $4463\pm4$ & (4.7) & $-0.05\pm0.002$ & $0.19\pm0.006$ \\
S05 & $4505\pm176$ &$4.71\pm0.96$ & $-0.18\pm0.19$ & ... \\ 
S08 & $4715\pm102$ & $4.39\pm0.28$ & $-0.32\pm0.03$  & ... \\ 
N09 & S08\tablenotemark{a} & S08 & S08 & $0.20\pm0.18$ \\ 
C11 & $4455\pm80$ & $4.67$ & ... & ...\\
A12 & S08 & S08 & S08 & $0.22\pm0.08$  \\
T13 & $4400\pm45$ & $4.36\pm0.1$ & $-0.26\pm0.14$ & ...\\
\cline{1-5}\\

\multicolumn{5}{c}{\textbf{HIP 13388 N=2 K1V}}\\
This work  & $4991\pm52$ & (4.6) & $-0.38\pm0.055$ & $0.26\pm0.037$ \\
C11 & $5095\pm 64$ & $4.59$ & $-0.15\pm0.1$ & $0.02$\\
S08 & $5040\pm48$& $4.39\pm0.08$& $-0.45\pm0.04$ & ... \\ 
N09 & S08 & S08 & S08 & $0.22\pm0.1$  \\ 
\cline{1-5}\\

\multicolumn{5}{c}{\textbf{HIP 10798 N=5 G8V}}\\
This work & $5312\pm19$ & (4.6) & $-0.54\pm0.0081$ & $0.14\pm0.008$ \\
V05 & $5374\pm44$ & $4.69\pm0.06$ & $-0.47\pm0.03$ & ...  \\
C11& $5481\pm80$ & $4.63$ & $-0.44\pm0.1$ & $0.17$ \\
\cline{1-5}\\

\multicolumn{5}{c}{\textbf{HIP 22278 N=1 G5V}}\\
This work & $5652 \pm68$ & (4.5) & $0.04 \pm0.056$ & $0.08\pm0.046$ \\
C11 &  $5721\pm 65$ &  $4.22$ & $0.13\pm0.1$ & $-0.01$ \\
\cline{1-5}\\

\multicolumn{5}{c}{\textbf{HIP 19589 N=1 G0V}}\\
This work & $5966\pm108$ &(4.5)& $-0.30\pm0.067$ & $0.15\pm0.057$\\ 
C11 &$5825\pm90$ & $3.75$ & $-0.17\pm0.1$ & $0.13$ \\
K13 & $5705\pm79$ & $3.40\pm0.15$ & $-0.52\pm0.1$ & $0.28\pm0.15$ \\  
\cline{1-5}\\

\multicolumn{5}{c}{\textbf{HIP 31415 N=1 F6V}}\\
This work &  $6295\pm108$ & (4.4) &$-0.55\pm0.067$ & $0.21\pm0.057$ \\
C11 & $6172\pm60$ & $3.94$ & $-0.31\pm0.1$ & $0.12$\\ 
\cline{1-5}\\

\multicolumn{5}{c}{\textbf{Sol N=909 G2V}}\\
This work & $5726\pm2 $ & (4.5) & $-0.03\pm 0.00$ & $0.01\pm 0.00$

\enddata
\tablenotetext{a}{Value reported is from S08}
\tablerefs{
\citetext{K13 \citealp{Kordopatis13}; C11 \citealp{Casagrande11}; 
N09 \citealp{Neves09}; S08 \citealp{Sousa08}; S05 \citealp{santos05}; 
V05 \citealp{Valenti05}; T13 \citealp{Tsantaki13}; A12 \citealp{Adibekyan12} }}
\tablecomments{[$\alpha$/Fe] values for N09 and A13 are the average of Mg, Ca, Si, and (Ti I + Ti II)/2. Note that
 [$\alpha$/Fe] values from C11 are not direct measurements and a measured by proxy from a statistical relation reported therein. 
Solar values are based on fits to $\sim900$ twilight spectra. 
Errors quoted for our stars are based on Table~\ref{stellarpropprectable}. Our log(g) values are those used during
fitting and should not be interpreted as a measurement. The values reported in this table 
do not include any adjustments for possible systematic errors.
}
\end{deluxetable}

\begin{deluxetable}{lcrrrr}
\tablewidth{0pc}
\tablecaption{Parameter Differences \label{stderrors}}
\tablehead
{
\colhead{Target} &
\colhead{Type} &
\colhead{$\Delta T_{eff}$ (K)} &
\colhead{$\Delta$[Fe/H] (dex)} &
\colhead{$\Delta$[$\alpha$/Fe] (dex)} 
}
\startdata
HIP 48331 & K5V &   $9\pm37$ 		& $+0.26\pm0.03$ 	& $-0.03\pm0.07$ \\ 
HIP 13388 & K1V &  	 $-69\pm53$ 		& $+0.03\pm0.05$ 	& $+0.08\pm0.09$ \\ 
HIP 10798 & G8V & 	 $-87\pm49$ 		& $-0.07\pm0.04$ 	& $+0.03\pm0.2$ \\
HIP 22278 & G5V & 	 $-69\pm94$ 		& $+0.17\pm0.11$ 	& $+0.09\pm0.20$ \\
HIP 19589 & G0V & 	 $209\pm124$ 	& $+0.05\pm0.10$ 	& $-0.08\pm0.13$ \\ 
HIP 31415 & F6V & 	 $123\pm124$ 	& $-0.24\pm0.12$ 	& $+0.09\pm0.21$ \\
\cline{1-5}\\
Average & & $-23\pm24$ & $+0.10\pm0.02$ & $+0.01\pm0.05$\\
\cline{1-5}\\
Twilights & G2V & $-51 \pm 2 $  & $-0.03\pm 0.002$ 	& $+0.01\pm 0.002$ \\
\enddata
\tablecomments{
Differences in our stellar parameters from the averages of the values reported in Table~\ref{stdparvals}. Deltas are Ours - Other. 
Twilight values are excluded from the average as the twilight spectra suffer from significantly higher scattered light 
and the small uncertainties would heavily bias the average. We note that the elevated [Fe/H] is driven entirely by HIP 48331: 
excluding it $\Delta$[Fe/H] becomes $-0.03\pm0.03$ dex, consistent with our twilight values. }
\end{deluxetable}

\subsection{RV Precision and Accuracy}

\subsubsection{Precision}
\label{rvprecisionsec}

Radial velocity variations of stars without a companion stem from one of five sources: 
(1) an inherent photon noise error ($\sigma_{phot}$) arising from the S/N and the number and shape of the stellar and telluric lines, 
(2) an instrumental error ($\sigma_{inst}$) based on the characteristics of M2FS spectra, 
(3) an error contribution due to our analysis ($\sigma_{anal}$), 
(4) intrinsic stellar variability ($\sigma_{stel}$) caused by stellar activity (e.g. stellar flares or star spots), and 
(5) variability in the bulk atmospheric motion along the line of sight that introduces a Doppler shift on our wavelength reference ($\sigma_{atm}$). 
We assume that all five sources add in quadrature to produce the observed dispersion ($\sigma_{obs}$), as follows:
\begin{displaymath}
\sigma_{obs}^{2}=\sigma_{phot}^{2}+\sigma_{inst}^{2}+\sigma_{anal}^{2}+\sigma_{stel}^{2}+\sigma_{atm}^{2}.
\end{displaymath}
Under this assumption, the observed velocity dispersion of a star with a known $\sigma_{stel}$ and observed under 
conditions with a known $\sigma_{atm}$ can be used to estimate the quadrature sum of the first three error terms, 
which we refer to as an effective measurement error, $\sigma_{meas}$,
\begin{displaymath}
\sigma_{meas}^{2}=\sigma_{phot}^{2}+\sigma_{inst}^{2}+\sigma_{anal}^{2}.
\end{displaymath}
Here we focus on estimating $\sigma_{meas}$ as a function of S/N, based on observations of the standard star HIP 48331.

\begin{figure}
\plotone{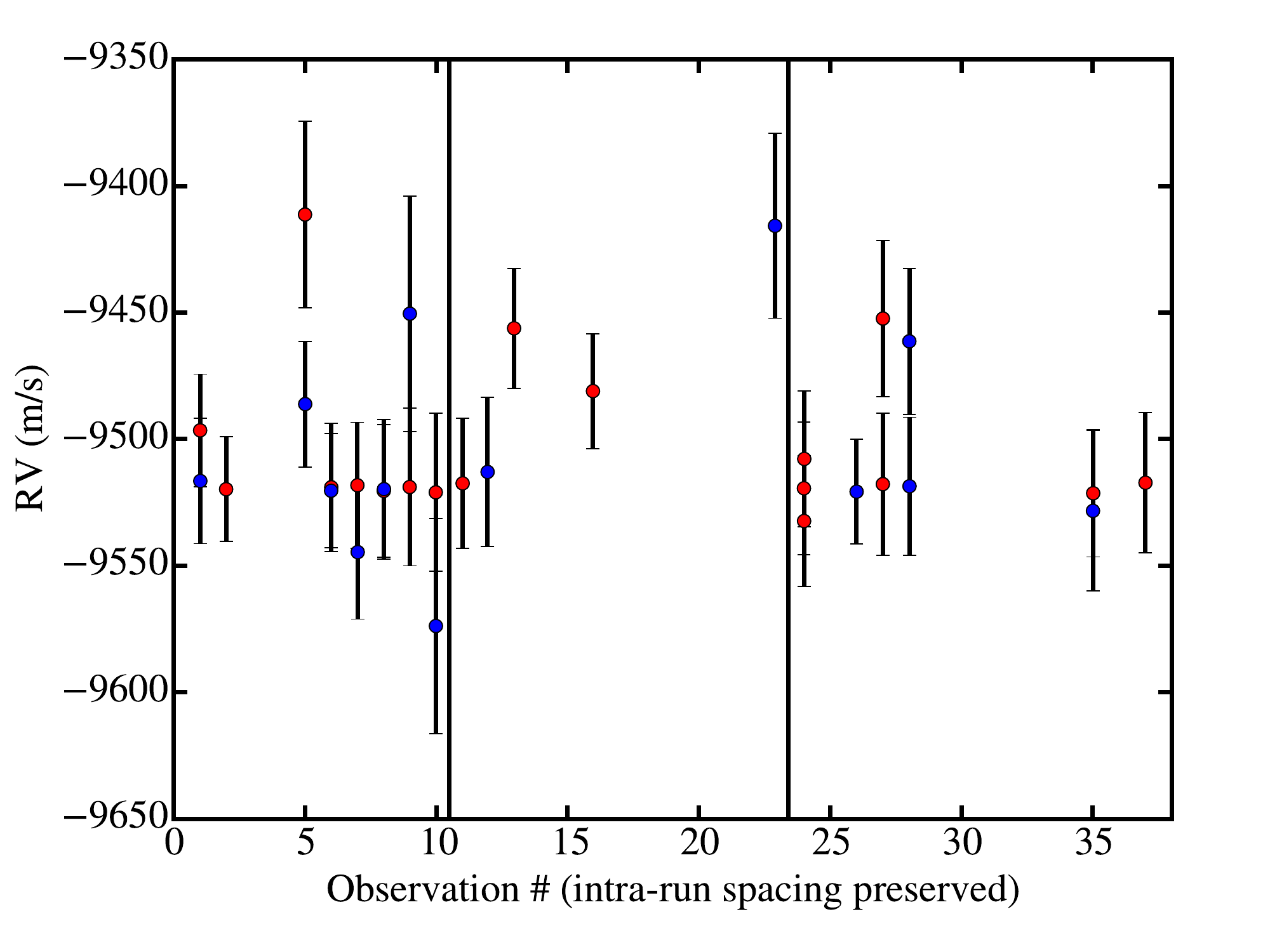}
\caption{\label{rvmeasures} A plot of our measurements of HIP 48331. Points are colored by the arm used for the observation, 
in this regard these observations represent a more 
stringent test of M2FS's stability than program stars which typically always use the same fiber and spectrographic channel.}
\end{figure}

We observed HIP 48331 35 times on 19 different nights; 9 nights have more than 1 epoch. Of these we use the 
31 spectra with S/N above 200 and R $>38,000$. Eighteen spectra were obtained using the red M2FS arm and 13 using the blue arm. The S/N of these spectra span between 200 and 300, with a median of 240. The resolving power of these spectra range from 40,000 to 64,000, with a median of 55,000 due to variations in the PSF width across the M2FS detector and soft instrument focus during early observing runs. The RV measurements of HIP 48331 are illustrated in Figure~\ref{rvmeasures}; these values have a standard deviation ($\sigma_{obs}$) of 23 m/s.

To simulate lower S/N spectra that are more representative of the open cluster stars surveyed here, we generated lower S/N versions 
of these 31 spectra and recomputed the best fit models and RVs from which new $\sigma_{obs}$ can be calculated. We generated the 
the lower S/N spectra by sampling a Poisson process at each pixel with expectation value of the measured electrons multiplied by 
the desired fractional reduction in mean S/N: e.~g.
\begin{displaymath}
x_i'=\rr{Poission}(sn' x_i/ sn)  
\end{displaymath}
where $x_i$ is the number of electrons measured at the $i^{th}$ pixel and prime denotes the new values. We also ensured that the 
simulated variance spectra included in an appropriate amount of Gaussian noise to include the effects of detector read noise. 
The resulting spectra have S/N levels of $\sim$150, 100, 80, 60, 50, 40, and 15. The spectra were then fit as described in 
Section~\ref{analysis}, treating each S/N level independently. This resulted in eight RV time-series (one for each S/N level) 
with each standard deviation yielding a measurement of $\sigma_{obs}$ at that S/N level. We also computed $\sigma_{phot}$ for each 
of the 248 spectra by applying the algorithm described in \citet{attaining3ms} to the telluric and stellar components of each best-fit 
model, adding the results in quadrature.

\begin{figure}
\plotone{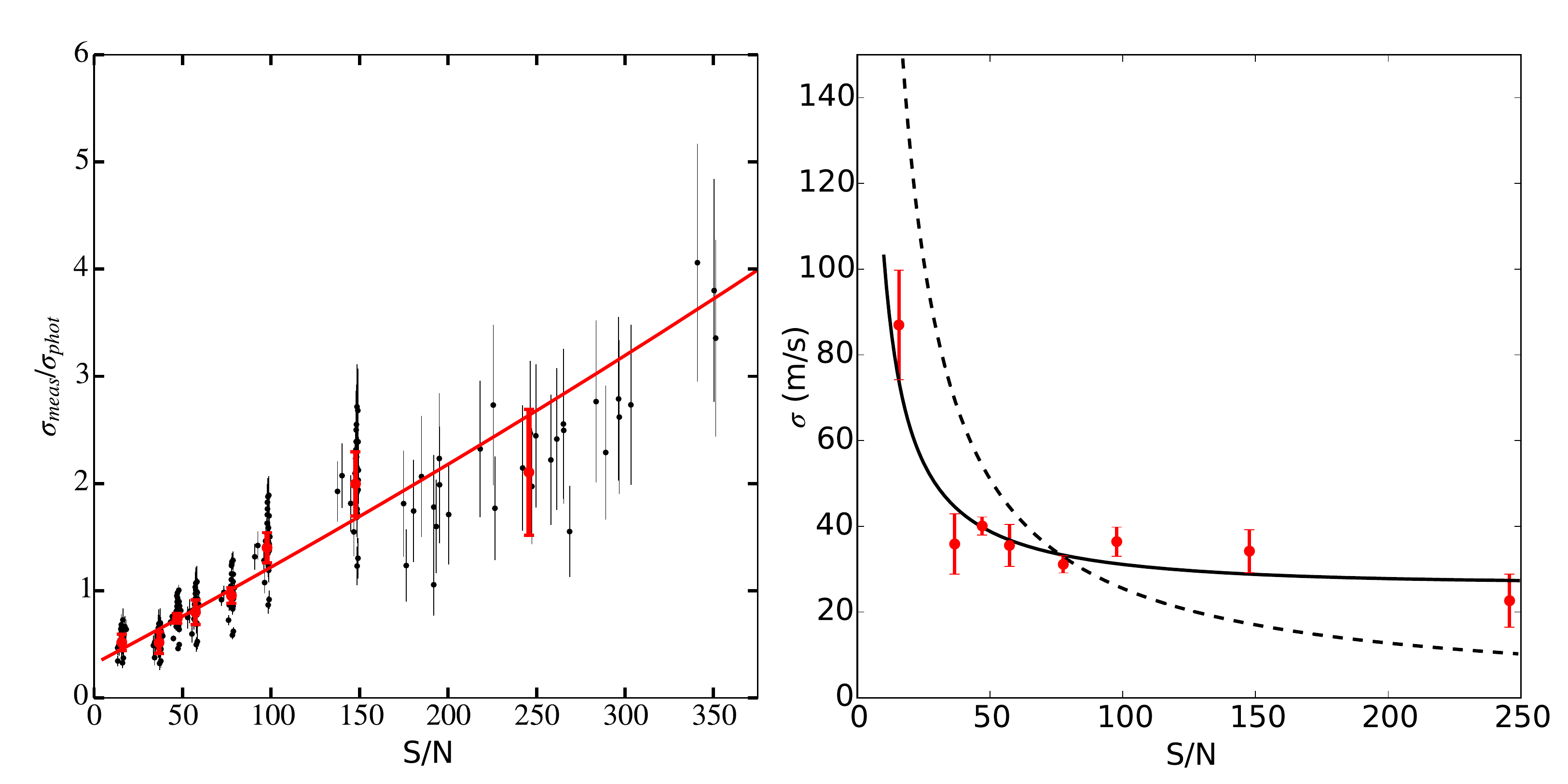}
\caption{\label{rvprecisionmeas} The left panel shows the ratio of $\sigma_{meas}$, computed from measurements of the RV 
standard at each resampled S/N step, to $\sigma_{phot}$, which is computed from the model of each spectrum. The individual points
are each of the RV standard spectra and are provided for visual reference. We fit to the means at each S/N bin. Errors are as described 
in the text. 
The right, most dispersed group of points reflects the native S/N of observations of HIP 48331. The right panel shows $\sigma_{meas}$
for each S/N bin along with an interpolated function generated using the fit in the left panel 
and the dashed curve. The dashed curve shows the mean of the $\sigma_{phot}$ values computed using the models from fits to the high-S/N
RV standard spectra with the calculation fed various S/N levels. This shows a clear indication 
that the algorithm overestimates uncertainty at low S/N. The plot also shows we are subject to a systematic floor of about 25 m/s
}
\end{figure}

\begin{figure}
\plotone{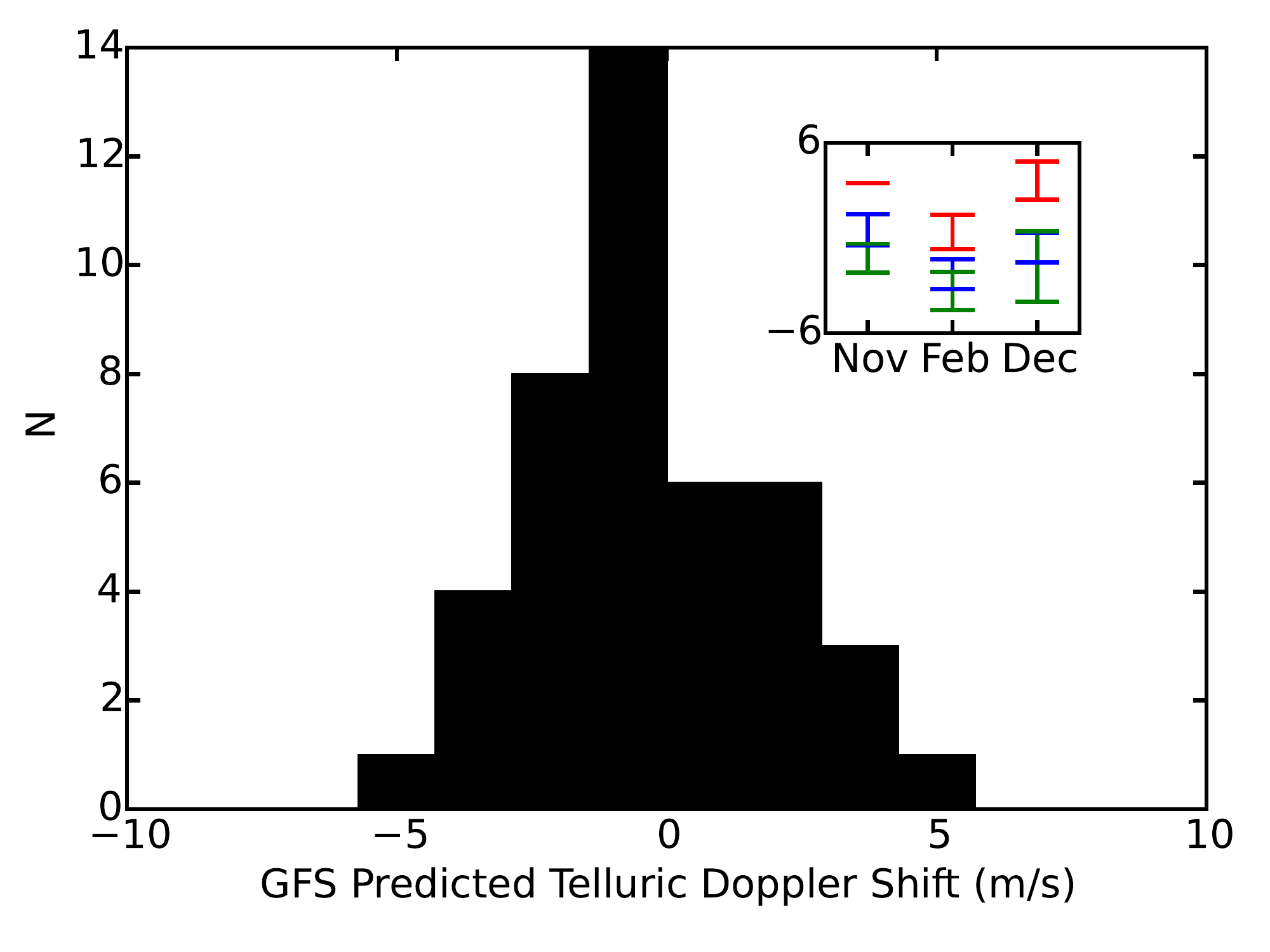}
\caption{\label{noaawindfig} A histogram of the telluric atmosphere imposed RV shifts to observations of HIP 48331, NGC~2516, 
and NGC~2422. The standard deviation about zero
is 2.3 m/s. The inset  shows the means and standard deviations from our November 2013, February 2014, and December 2014 
observing runs for the three sets of targets separately in green, blue, and red respectively.}
\end{figure}

To obtain $\sigma_{meas}$ from the eight $\sigma_{obs}$ values calculated above, we subtracted off a stellar variability of 
$\sigma_{stel}=5.0$ m/s \citep{gaiarv} and an atmospheric variability of $\sigma_{atm}=2.5$ m/s (determined as shown later in this 
section, see also Figure~\ref{noaawindfig}). 
These values can be compared directly to the the mean $\sigma_{phot}$ values for each of the eight S/N bins. Oddly, we found that
would result in imaginary errors below a S/N of $\sim60$. In Figure~\ref{rvprecisionmeas} we plot both $\sigma_{meas}$ and 
$\sigma_{phot}$, which shows that we measure our RVs with greater precision that anticipated at low S/N. 
In Figure~\ref{rvprecisionmeas}, we also show the ratio of  $\sigma_{meas}$ to the mean of $\sigma_{phot}$ at each S/N bin. As an additional 
reference we also plot the ratio of each bin $\sigma_{meas}$ to each of the $\sigma_{phot}$ in that bin. This suggests an approximately 
linear relation between our measurement error and the  $\sigma_{phot}$ value we computed for each spectrum. We adopted errors for the 
ratio from two sources: (1) the standard deviation of $\sigma_{phot}$ in each S/N bin contributes directly and (2) 
an estimate of the error in $\sigma_{obs}$ that was obtained by computing our best-fit models with a small number of 
slightly perturbed initial RVs for each spectrum in each bin, adopting the standard deviations of the resulting $\sigma_{obs}$ 
values as an uncertainty on $\sigma_{meas}$ in each S/N bin. 

We fit the ratio of $\sigma_{meas}$ to the mean of the $\sigma_{phot}$ for each S/N bin and use the result as a scaling relation 
to convert $\sigma_{phot}$ to $\sigma_{meas}$ provided a S/N. This technique allows us to account for some, if not all, 
of the increased uncertainty in spectra that are at a lower resolution (e.g. due to mis-focus) than the typical RV standard 
observation, are of more rapidly rotating stars, or otherwise possess a different number or strength of stellar lines. 
The errors in Figure~\ref{rvmeasures} have been scaled in this manner. 

\begin{figure}
\plotone{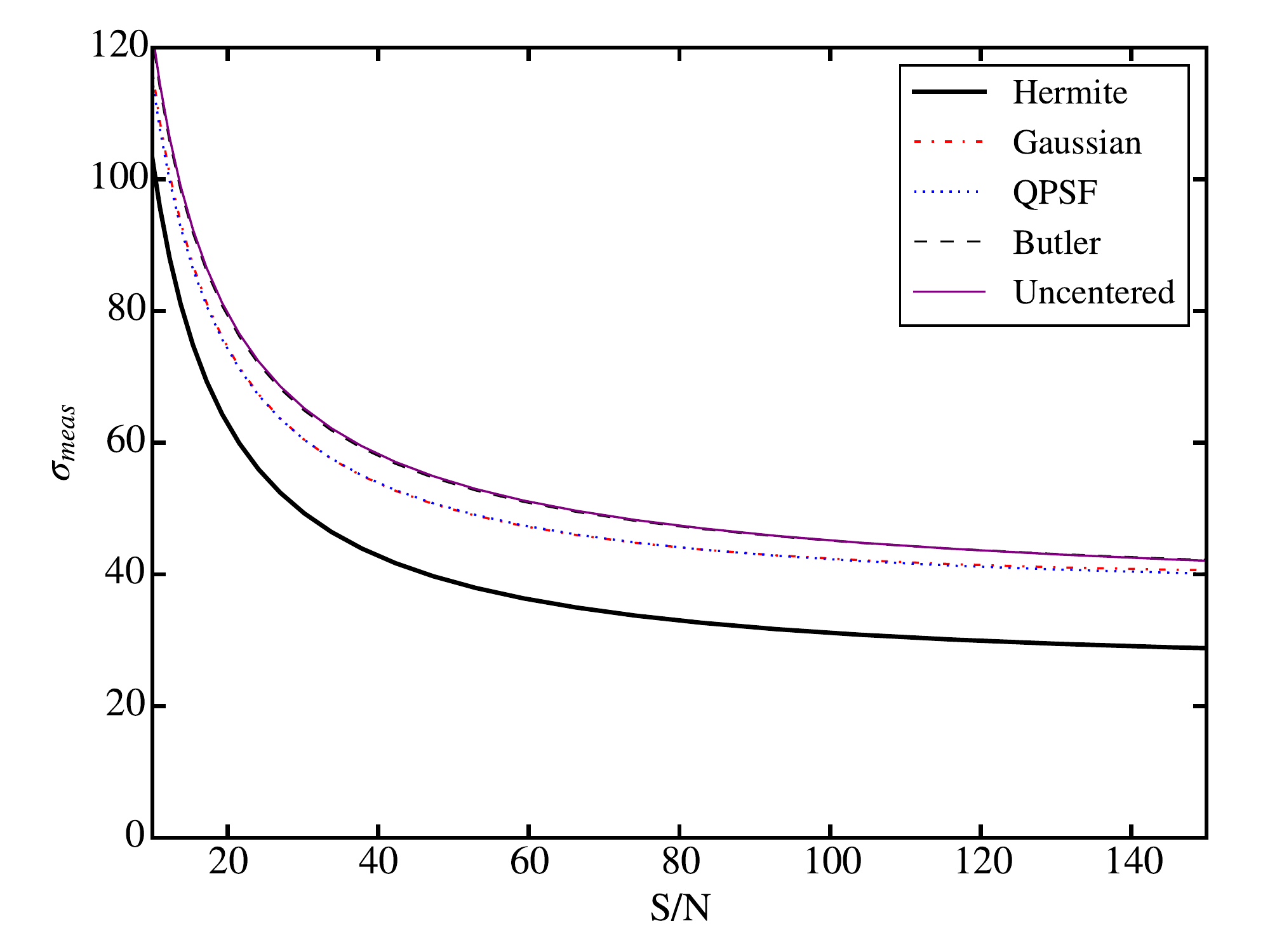}
\caption{\label{rvprecisionpsf} This plot shows the impact various PSF modeling choices have on $\sigma_{meas}$.
The solid black line denotes $\sigma_{meas}$ for our adopted analysis technique (c.f Fig \ref{rvprecisionmeas}).  The red
dash-dotted line corresponds to our analysis but using a simple, fixed Gaussian PSF. The blue dotted line -- essentially on top of the
red line --  is for fits done using a Gaussian PSF with a FWHM as described by a quadratic. The thin, dashed black line and the thin 
purple line -- also nearly superimposed -- correspond to fits done with the PSF prescription of  \citet{attaining3ms} and our adopted,
Hermite prescription but without the enclosed power constrained to the central pixel. }
\end{figure}

\begin{figure}
\plotone{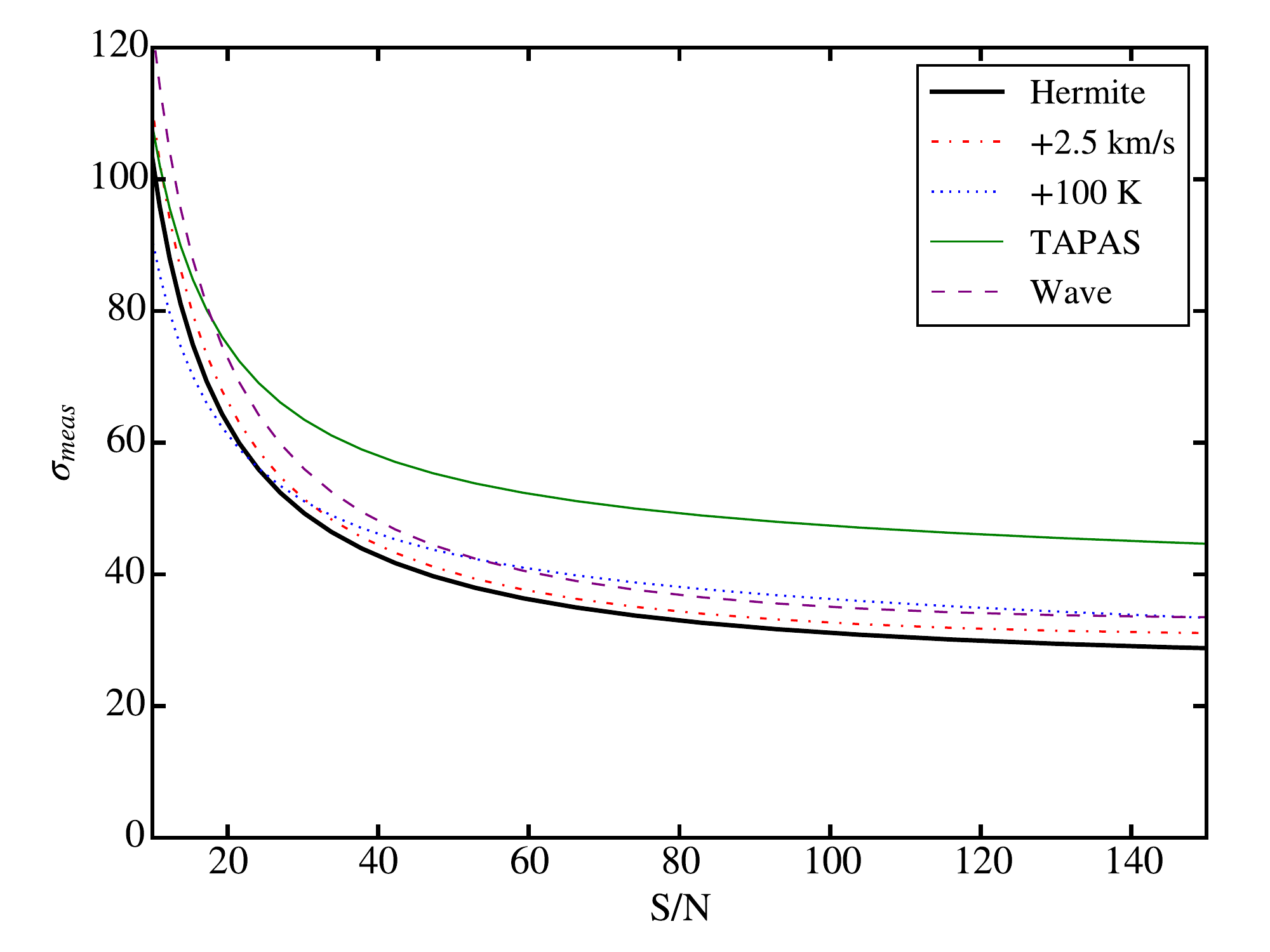}
\caption{\label{rvprecisionspt}
This plot shows the impact of various factors on our attained RV measurement precision. The solid black line denotes $\sigma_{meas}$ 
for our adopted analysis technique (c.f Fig \ref{rvprecisionmeas}). The red dash-dotted line corresponds to a fits done with a $v_r\sin(i)$
2.5 km/s larger than the optimal value. The blue dotted line corresponds to fits done with $T_{eff}$ forced 100 K above our adopted value.
The purple dashed line is for fits done without holding the $4^{th}$ and higher order wavelength parameters fixed as described in Section~\ref{modelfittingsec}.
Finally, the solid green line represents our results when we use the TAPAS synthetic telluric spectra as our wavelength reference instead of the NSO
empiric spectrum.}
\end{figure}

\begin{figure*}
\centering\includegraphics[width=1.2\textwidth,clip=true,trim=2cm 0cm 1cm 0cm]{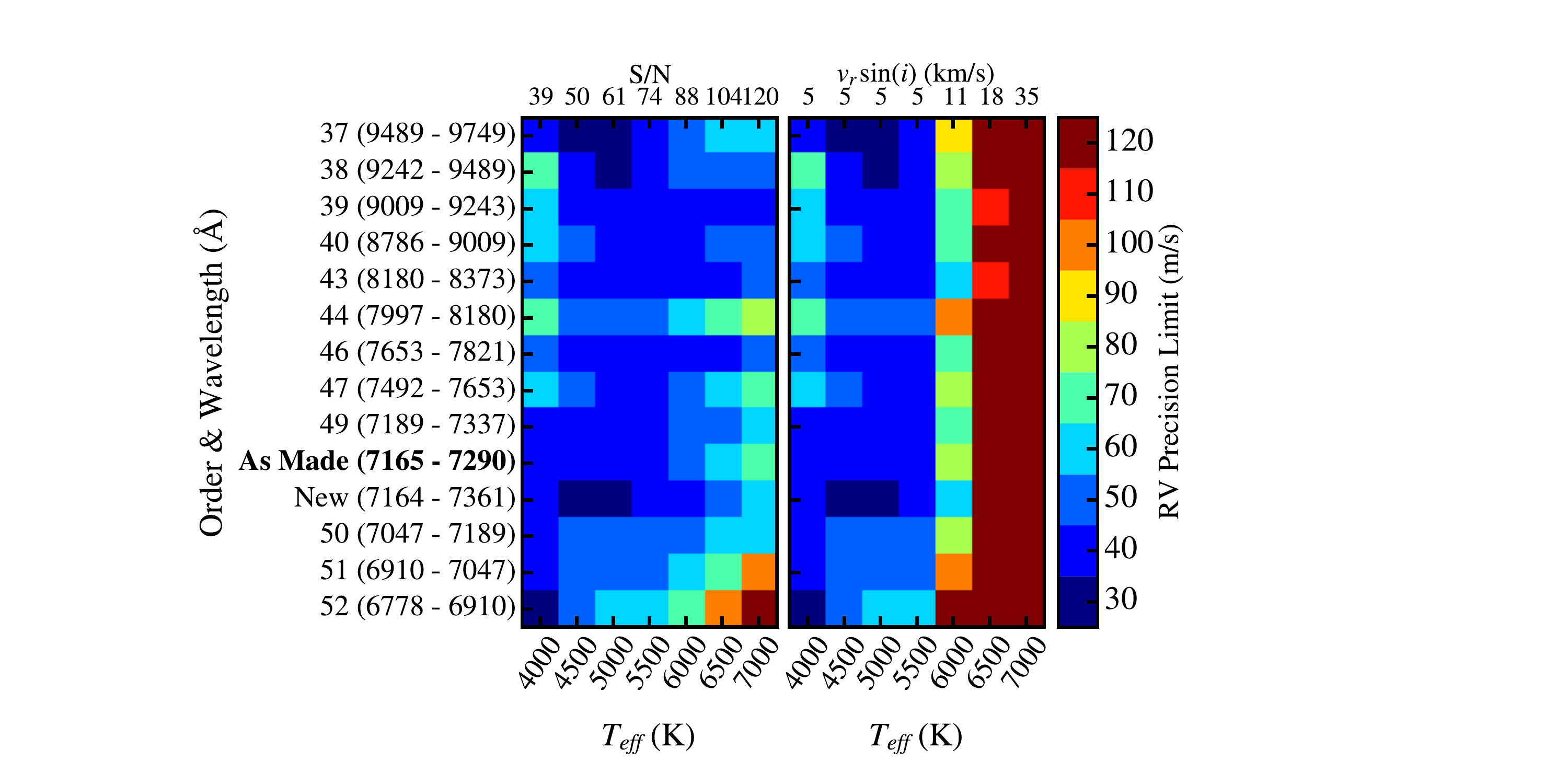}
\caption{\label{rvprecision2}  
This is an updated version of Figure~\ref{precisionfig} where the algorithmic uncertainties have been corrected 
as described in Section~\ref{rvprecisionsec}. Both plots are at our median observed resolving power of 50,000. The S/N of each 
spectrum is determined by assuming equidistant MS dwarfs where a S/N of 50 is attained for a K5 dwarf. 
The resulting scale is marked at the top of the left plot, which assumes all stars have a $v_r\sin(i)$ of 5 km/s. The right 
plot uses the same S/N scale combined with the median $v_r\sin(i)$ values we measure for stars with the stated $T_{eff}$ in NGC~2516 
and NGC~2422. The right panel thus presents a worst case scenario for our technique as these clusters are some of the youngest 
suitable for precision RV work. In addition to standard M2FS orders we note the truncated order 49 used in this paper as ``As Made'' 
and our expectations for the new filter described in Section~\ref{filterissuesec} as ``New.''
}
\end{figure*}

From this analysis we find M2FS has a limiting RV 
precision of about 25 m/s, though the $\sigma_{phot}$ values we computed at high S/N suggest an
additional 15 m/s precision gain may be possible at higher S/N ratios. A potential culprit in our modeling process is as yet unclear. 
In Figures \ref{rvprecisionpsf} and \ref{rvprecisionspt} we use the same process to show the impact various modifications to 
our analysis have on achievable precision, some of which are discussed in further detail in the following paragraphs. 
Finally, Figure~\ref{rvprecision2} shows an updated version of Figure~\ref{precisionfig} with the corrections discussed above.

\paragraph{Atmospheric Variability}
We estimated the impact bulk atmospheric motions have on our wavelength reference by integrating the water vapor weighted wind 
speed along the line of sight using data from the NOAA GFS forecast models \citep{noaagfs}. Using the GFS model closest in time to our 
data the forecast is within 3 hours of the model's initial conditions. These models have an RMSE wind vector error
of about 3 m/s three days (!) in the future. Perturbing the integrals by this error has a maximum impact of about 1 m/s, 
with typical values less than a tenth of that. The resulting contributions for our data on HIP 48331, NGC~2516, and NGC~2422 are 
shown in Figure~\ref{noaawindfig}. While pointing directly into or along the jet-stream would exhibit a clear signature 
at the $\pm5$ m/s level, typical values are not particularly significant to our efforts. As mentioned, we adopt $\sigma_{atm}=2.5$ m/s.

\paragraph{Sky Emission}
Ideally the spectra we obtained of our RV standard would be completely representative of our program stars. Our cluster targets
are, however, significantly fainter than our RV standard and so many of them exhibit a number of strong sky emission lines 
(c.f. Figure~\ref{exampleextractions} and \ref{fitexamplefig}). 
To better asses their impact on our RV precision we took a subset of 60 spectra and extracted and fit the spectra from images prior to stacking. 
This gave us a sample of 4 or 5 RVs from spectra obtained consecutively. These spectra spanned a S/N of about 15 -- 45 for targets of 
spectral type $\sim$ K3 - F5. The $\sigma_{obs}$ for these RVs was in agreement with that expected based on our  
$\sigma_{meas}$ relation. We see some evidence that our better-than-anticipated RV precision stems from our use of the mean RV as a prior.
If fit with an initial RV far from the multi-epoch mean we observe an increased $\sigma_{obs}$ at low S/N, though still somewhat below that 
predicted by \citet[][c.f. Figure~\ref{rvprecisionmeas}]{attaining3ms}.

\paragraph{PSF effects}
We found very little difference in the results of the multi-Gaussian parameterization of \citet{attaining3ms} and a Gauss-Hermite kernel
when the center of its enclosed power is not constrained to the central pixel undergoing convolution. 
The latter is faster with many fewer parameters and once the enclosed power 
is constrained to the central sub-pixel we find it exhibits enhanced stability (c.f. Figure~\ref{rvprecisionpsf}). Both yielded slightly worse 
performance than a simple Gaussian and we did not find an improvement in RV precision by using a variable, but symmetric PSF. 
It may be worth investigating a hybrid approach where the components of the Hermite parametrization are allowed to vary with pixel. 

\paragraph{Model Spectra}
As an additional test on RV precision and the impact our use of the PHOENIX grid has we model the twilight spectra both with the 
PHOENIX grid and using the empiric Solar spectrum of \citet{kurucz} as the template. We select the $\sim$ 600 twilight spectra in images with 
mean S/N above 100 (100 -- 650, mean of 320). For these spectra we measured a $1\sigma$ RV scatter within each twilight image of
$23\pm1.4$ m/s when fitting with the PHOENIX models and $28\pm2.3$ m/s using Kurucz's empiric Solar spectra. This strongly 
suggests that the PHOENIX templates are not limiting our RV precision. We see evidence of a slight quadratic dependence of the measured 
RV on the spectrum's CCD position. This suggests that the RV zero point and wavelength zero points are slightly affecting our 
dispersion, though we note that program stars are typically observed in the same fiber. Fitting and removing this effect reduced 
the scatter to $18\pm1.2$ m/s and $23\pm2.3$ m/s, respectively.

\subsubsection{Accuracy}
We estimated the accuracy of our RVs by looking at the differences between our values and those in the literature for each 
of our six standard stars. We report these differences in Table~\ref{stdrverrors} and find an offset of $74\pm72$~m/s from the scale 
of \citet{gaiarv}, albeit with significant scatter.  We also saw a slight indication that RVs measured in our lowest S/N bin are slightly 
shifted relative to the higher S/N bins by $27\pm17$ m/s.

\begin{deluxetable}{lccr}
\tablewidth{0pc}
\tablecaption{Standard RV Differences \label{stdrverrors}}
\tablehead
{&&&
\colhead{$\Delta$RV}
\\
\colhead{Target} &
\colhead{Type} &
\colhead{N} &
\colhead{(m/s)}
}
\startdata
HIP 48331 & K5V & 30 & $2\pm7$ \\
HIP 13388 & K1V & 2 & $41\pm40$ \\
HIP 10798 & G8V & 5 & $139\pm7$ \\
HIP 22278 & G5V & 3 & $173\pm17$ \\
HIP 19589 & G0V & 1 & $-77\pm94$ \\
HIP 31415 & F6V  & 1 & $173\pm78$ \\
\cline{1-4}\\
Average &  &  & $74\pm72$
\enddata
\tablecomments{Differences in our RVs compared to the values reported in Table~\ref{stdtargtable}. Differences are Ours - Other.
}
\end{deluxetable}

\section{Discussion}
\label{discussionsec}

Given M2FS's unique ability to efficiently obtain precision optical spectra capable of determining accurate stellar properties and 
precision RVs, here we briefly summarize the broader scientific impact this instrument could have for both finding 
exoplanets in open clusters and improving our understanding of the stars in these clusters.

\subsection{M2FS as a Tool for Finding Planets in Open Clusters}

\begin{figure}
\plotone{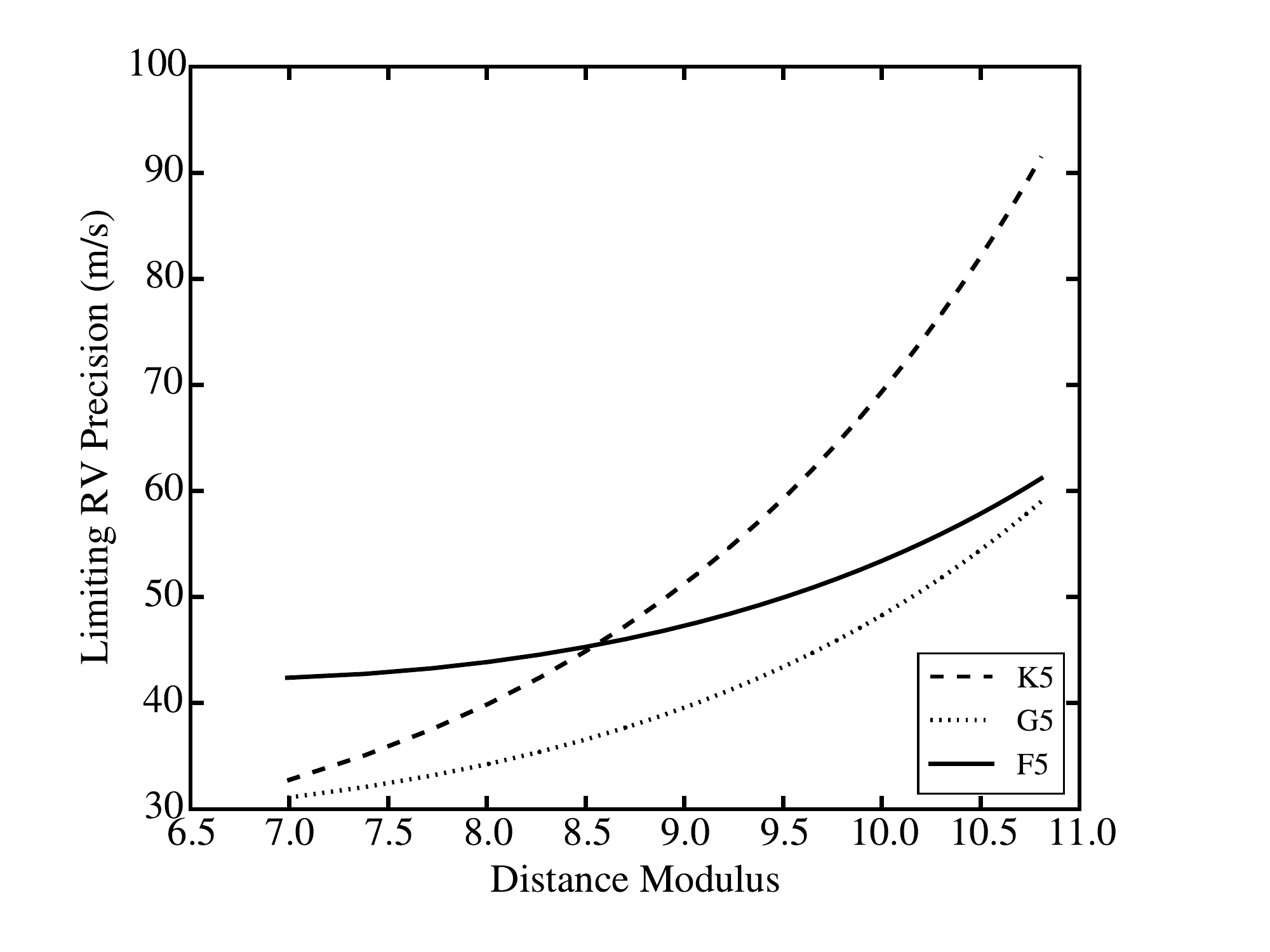}
\caption{\label{precisionDM} A plot of anticipated RV precision for quiescent, slowly-rotating K5, G5, and F5 stars as function of distance modulus after 2.5 hours observing in one arc second seeing with the corrected filter.}
\end{figure}

Though once unexpected, it is now clear that a great many hot-gas giants exist. Assuming 1.2\% of F5-K5 stars in open
clusters harbor hot gas planets (P $< 10$ days, $M\sin(i)>0.1\rr{M_{Jup}}$) \citep{wright12,meibom13} and 
given our achieved precision we can predict the limits of our technique. For example, we expect M2FS will
be capable of attaining a S/N 25 spectrum of a K5V star at a DM of 9.5 in 4 hours (effectively 2 minutes per star). This would be sufficient for an RV precision of ~$\sim55$ m/s, with brighter members increasingly limited by the systematics in our analysis. In this hypothetical cluster we would then be sensitive to $\sim75$\% of known hot gas giants. Figure~\ref{precisionDM} shows our anticipated RV measurement precision as a function of distance modulus using the new filter. This implies we could reasonably expect one Hot-Jupiter per M2FS pointing, provided targets are available for the majority of fibers. 

Table~\ref{futuretargtable} lists the eight nearby open clusters that matched the cluster selection criteria given in Section~\ref{targselecsec}. From these clusters we can obtain $\sim15$ M2FS pointings and would expect about as many exoplanet candidates. We highlight that these clusters span a range of ages and are thus well suited to help build a sample of exoplanets that addresses the formation and migration issues discussed in the introduction. 

\begin{figure}
\plotone{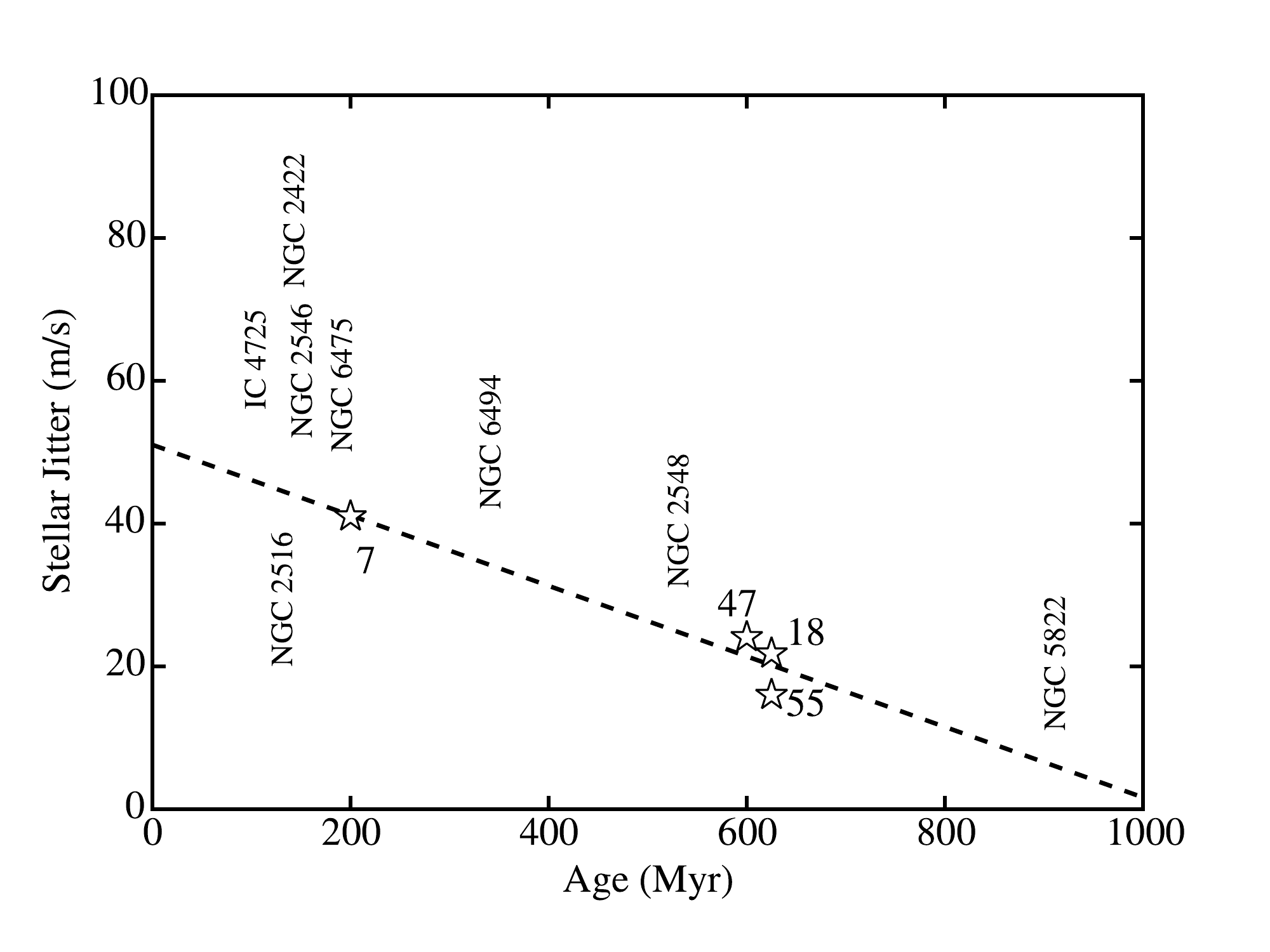}
\caption{\label{agejitter} A plot of the literature values for stellar jitter as a function of age. Sample sizes are noted by each point. 
The dashed line is a simple, linear fit to guide the eye. The various clusters well matched to M2FS's grasp are noted at their
approximate ages. Data is from \citet{paulsonyelda,paulson04,quinn12,quinn14}.
}
\end{figure}

This program would also well characterize the Doppler uncertainty commonly referred to as stellar jitter as a function of age. This issue is not yet well constrained \citep[c.f.][]{Mahmud11,Lagrange13} and the present state of our knowledge is largely summarized in Figure~\ref{agejitter}, which shows the stellar jitter as a function of age for a young association and two open clusters along with the ages of open clusters our expanded survey will target. Extant data is far from homogenous -- a mix of both optical and infrared spectroscopy -- and suffers from rather small sample sizes. By surveying a large number of stars in clusters over a range of ages we will well constrain this effect with age and determine what truly constitutes ``too young'' to measure precision RVs.

We note that M2FS's strength is in identification: though the ability to survey large numbers of stars at this precision is unmatched, we would suggest that followup of promising candidates is better suited to traditional monitoring programs.

\subsection{M2FS as a Tool for Studying Open Cluster Stellar Populations}

The success of any large scale RV survey of open clusters for planets depends critically on having a carefully vetted sample 
to survey. M2FS is poised to do this.  First, precision RVs will help confirm membership, especially when combined with Gaia 
proper motions. Second, precision RVs can identify spectroscopic binaries that are typically poor targets for exoplanet searches. Finally, the high 
dispersion spectra allow measures of $v_r\sin(i)$, $T_{eff}$, and metallicity, with the latter two yielding stellar masses 
from evolutionary models. When coupled with photometric periods determined by LSST, we could even determine stellar 
inclinations and identify edge-on systems. The flexible nature of M2FS also means that we could also obtain spectra of the 
Ca~II H and K region for all of our targets with a relatively modest overhead ($\sim20$ min total for targets in NGC~2516 and NGC~2422), 
further helping calibrate stellar ages.

A number of recent papers have also highlighted the importance both stellar multiplicity and metallicity play in star formation and have
drawn attention to various gaps in current simulations of cluster evolution \citep{paunzen10,geller10,duchene07}.
Our stellar properties directly address such gaps by characterizing the cluster chemical environment while our RVs allow
robust identification of binaries and brown dwarfs (constrained by our time baseline), characterizing the kinematic environment. 
Such a dataset can contribute to the initial conditions used in dynamical simulations of cluster formation. 

We also note there is an absence of precision internal kinematics for open clusters. This dataset is useful to study 
the internal dynamics of open clusters at the 10 m/s level. With an anticipated precision of better than 10 $\rr{\mu}$as/yr ($~\sim$ 20 m/s at 500 pc) 
\citep{lindegren10,lindegren12}, once GAIA data is available for our targets the combined dataset will offer an unprecedented 
3D kinematic picture of stars within open clusters, providing a useful tool to study internal kinematics.

\section{Summary}

We have presented a program to use the Michigan/Magellan Fiber System to obtain multiplexed spectroscopy of solar analog 
stars in nearby ($<$1 kpc) open clusters with the intent of identifying exoplanet host stars for subsequent followup.  Our 
technique uses telluric lines in the 7230 \AA\ region as a wavelength reference and is presently capable of measuring RVs 
with a precision of 25-60 m/s, depending on S/N. We also obtain precise and accurate measurements of $T_{eff}$, [Fe/H], 
[$\alpha$/Fe], and $v_r\sin(i)$ for all of our target stars thereby enabling characterization of the cluster environment. 
This paper is the first in a series of papers on our efforts and described 
our analysis procedure in detail. The next paper in this series will report the RVs and stellar parameters of targets in NGC~2422 
and NGC~2516 and carry out an analysis of the companion detectability therein. Later papers will use our sample to study 
binaries in our sample and analyze emergent cluster properties.

\begin{deluxetable}{lcccrc}
\tablewidth{0pc}
\tablecaption{Potential Target Clusters\label{futuretargtable}}
\tablehead
{

& 
\colhead{Age} & 
\colhead{Distance} &
\colhead{$R_{Cen}$} & & 
\\
\colhead{Cluster} & 
\colhead{(Myr)} & 
\colhead{(pc)} &
\colhead{(deg)} & 
\colhead{[Fe/H]} & 
\colhead{$\sim N_{Targ}$}
}
\startdata

IC 4725 (M25) &  93  & 560 & 0.5  & -0.3 & 100   \\
NGC 2546 & 140 & 930 & 0.8 & +0.12 & 200 \\
NGC 6475 (M 7)   & 180 & 300 & 0.8 & +0.14 & 150 \\
NGC 6494 & 330 & 650 & 0.5 & +0.09 & 200 \\
NGC 2548 	     &  520 & 790 & 1.1 & +0.08 & 100 \\
NGC 5822 & 900 & 800 & 0.5 & -0.02 & 500  \\
NGC~2422 (M 47) & 132 & 491 & 0.3 & +0.02 & 160 \\
NGC~2516 	     & 120 & 346 & 0.4 & -0.18 & 330 \\
\enddata

\tablecomments{The numbers given for NGC~2516 and NGC~2422 are in addition to those already observed. $R_{Cen}$ refers to the
 approximate half-light radius of the clusters, while $N_{Targ}$ is the approximate number of members available near the cluster 
 centers within the field of view of M2FS.
}
\end{deluxetable}

\acknowledgments
The authors gratefully acknowledge valuable discussions with Sam Quinn, Peter Gao, Justin Cantrell, and Colin Slater. J.B. and M.M. acknowledge support from NSF AAG grant 1312997 and the NFS/MRI development grant 0923160. R.W. acknowledges support from NSF AAG grant 1009634 and NASA Origins of Solar Systems grant NNX11AC32G. PyRAF is a product of the Space Telescope Science Institute, which is operated by AURA for NASA. This research made use of Astropy, a community-developed core Python package for Astronomy (Astropy Collaboration, 2013).

Facilities: \facility{Magellan:Clay(M2FS)}

\end{document}